\documentstyle[12pt,aasms4]{article}

\newcommand{\dotprod}{{{\scriptscriptstyle \stackrel{\bullet}{{}}}}}

\def\lbar{\mathrel{\rlap{
\raise3pt\hbox{\hskip-1.75pt-}}
\hbox{$\lambda_e$}}}

\def\simge{%
\mathrel{\rlap{\raise 0.511ex
\hbox{$>$}}{\lower 0.511ex \hbox{$\sim$}}}}

\def\simle{%
\mathrel{\rlap{raise 0.511ex
\hbox{$<$}}{\lower 0.511ex \hbox{$\sim$}}}}

\def\ni{\noindent}
\begin{document}

\newpage

\title{THE EQUATION OF STATE OF NEUTRON-STAR MATTER IN STRONG MAGNETIC FIELDS}
\author{A. BRODERICK, M. PRAKASH, AND J.M. LATTIMER}

\affil{Department of Physics and Astronomy, State University of New
York at Stony Brook \\ Stony Brook, NY 11974-3800}

\begin{abstract}

We study the effects of very strong magnetic fields on the equation of
state (EOS) in multicomponent, interacting matter by developing a
covariant description for the inclusion of the anomalous magnetic
moments of nucleons. For the description of neutron star matter, we
employ a field-theoretical approach which permits the study of several
models which differ in their behavior at high density. Effects of
Landau quantization in ultra-strong magnetic fields ($B>10^{14}$
Gauss) lead to a reduction in the electron chemical potential and a
substantial increase in the proton fraction.  We find the generic
result for $B>10^{18}$ Gauss that the softening of the EOS caused by
Landau quantization is overwhelmed by stiffening due to the
incorporation of the anomalous magnetic moments of the nucleons.  In
addition, the neutrons become completely spin polarized.  The
inclusion of ultra-strong magnetic fields leads to a dramatic increase
in the proton fraction, with consequences for the direct Urca process
and neutron star cooling.  The magnetization of the matter never
appears to become very large, as the value of $|H/B|$ never deviates
from unity by more than a few percent.  Our findings have implications
for the structure of neutron stars in the presence of large frozen-in
magnetic fields.

\ni {\em Subject headings:} stars: neutron stars -- equation of state --
stars: magnetic fields
\end{abstract}

\newpage
\section{INTRODUCTION}

Recent observational and theoretical studies motivate the
investigation of the effects of ultra-strong magnetic fields
($B>10^{14}$ Gauss) on neutron stars.  Several independent
arguments link the class of soft $\gamma-$ray repeaters
and perhaps certain anomalous X-ray pulsars with neutron stars having
ultra strong magnetic fields -- the so-called magnetars (Paczy\'nski
1992; Thompson \& Duncan 1995, 1996; Melatos 1999).  In addition, two
of the four known soft $\gamma-$ray repeaters directly imply, from
their periods and spin-down rates, surface fields in the range $2-8\times
10^{14}$ Gauss.  Kouveliotou et al. (1998, 1999) argue from the
 population statistics of soft $\gamma-$ray repeaters that magnetars
constitute about 10\% of the neutron star population.  While some
observed white dwarfs have large enough fields to give ultra-strong
neutron star magnetic fields through flux conservaton, it does not
appear likely that such isolated examples could account for a
significant fraction of ultra-strong field neutron stars.  Therefore,
an alternative mechanism seems necessary for the creation ultra-strong
magnetic fields in neutron stars.  Duncan \& Thompson (1992, 1996) suggested
that large fields (up to $3\times10^{17}\times (1\rm{~ms}/P_i)$ Gauss, where
$P_i$ is the initial rotation period) can be generated in nascent
neutron stars through the smoothing of differential rotation and
convection.

These developments raise the intriguing questions:\\
(1) What is the largest  frozen-in magnetic field
a stationary neutron star can sustain?,
and, \\
(2) What is the effect of such ultra-strong magnetic fields on the maximum
neutron star mass?

The answers to both of these questions hinge upon the effects strong
magnetic fields have both on the equation of state (EOS) of
neutron-star matter and on the structure of neutron stars.  In this
paper, we will focus on the effects of strong magnetic fields on the
EOS.  Subsequent work will be devoted to investigating the effects of
strong fields on the structure of neutron stars, incorporating the
EOSs developed in this work.

The magnitude of the magnetic field strength $B$ needed to dramatically
affect neutron star structure directly can be estimated with a
dimensional analysis (Lai \& Shapiro 1991) equating the magnetic field
energy $ E_b \sim (4\pi R^3/3)(B^2/ 8\pi) $ with the gravitational
binding energy $ E_{B.E.}  \sim GM^2/R$, yielding $B \sim 2\times
10^{18}\left(M/ 1.4 {\rm M}_\odot\right) \left(R/ 10\ {\rm km}\right)^{-2}$
Gauss, where $M$ and $R$ are, respectively, the neutron star mass and
radius.

The magnitude of $B$ required to directly influence the EOS can be
estimated by considering its effects on charged particles.  Charge
neutral, beta-equilibrated, neutron-star matter contains both
negatively charged leptons (electrons and muons) and positively
charged protons.  Magnetic fields quantize the orbital motion (Landau
quantization) of these charged particles.  Relativistic effects become
important when the particle's cyclotron energy $e\hbar~B/(mc)$  is
comparable to it's mass (times $c^2$). The magnitudes of the so-called
critical fields are $B_c^e= (\hbar c/ e) \lbar^{-2} =
4.414\times10^{13}$ Gauss and $B_c^p = (m_p/m_e)^2 B_c^e =
1.487\times10^{20}$ Gauss for the electron and proton, respectively  (
$\lbar=\hbar/m_ec\simeq 386$ fm is the Compton wavelength of the
electron).  It will be convenient to measure the field strength $B$ in
units of $B_e^c$, {\it viz.}, $B^*\equiv B/B^c_e$.  When the Fermi
energy of the proton becomes significantly affected by the magnetic
field, the composition of matter in beta equilibrum is significantly
affected.  In turn, the pressure of matter is significantly affected.
We show that this occurs when $B^*\sim10^5$, and will lead to a
general softening of the EOS.

In neutron stars, magnetic fields may well vary in strength from the
core to the surface.  The scale lengths of such variations are,
however, usually much larger than the microscopic magnetic scale
$l_m$, which depends on the magnitude of $B$.  For low fields, for
which the quasi-classical approximation holds, $l_m \simeq
(\lbar^2/B^*)(3\pi^2n_e)^{1/3}\approx 10^5(n_e/n_s)^{1/3}/B^*$ fm,
where $n_e$ is the number density of electrons and $n_s$ is the normal nuclear
saturation density (about 0.16 fm$^{-3}$).  For high fields, when only
a few Landau levels are occupied, $l_m \simeq 2\pi^2n_e
(\lbar^2/B^*)^2\approx7\times10^9(n_e/n_s)/B^{*2}$ fm.  In either case,
the requirement that $R >> l_m$ is amply satisfied; hence, the magnetic
field $B$ may be assumed to be locally constant and uniform as far as
effects on the EOS are concerned.

In non-magnetic neutron stars, the pressure of matter ranges from
$2-5~{\rm MeV~fm}^{-3}$ at nuclear density to $200-600~{\rm
MeV~fm}^{-3}$ at the central density of the maximum mass
configuration, depending on the EOS (Prakash et al. 1997).  These
values may be contrasted with the energy density and pressure from the
electromagnetic field: $\varepsilon_f = P_f = B^2/(8\pi) = 4.814
\times 10^{-8} B^{*~2}~{\rm MeV~fm}^{-3}$.  The field contributions
can dominate the matter pressure for $B^* > 10^4$ at nuclear densities
and for $B^* > 10^5$ at the central densities of neutron stars, and
must therefore be included whenever the field dramatically influences the
star's composition and matter pressure.

In strong magnetic fields, contributions from the anomalous magnetic
moments of the nucleons must also be considered.  Experimentally,
$\kappa_p = \mu_N \left(g_p/2 - 1 \right)$ for the proton, $\kappa_n =
\mu_N g_n/2$ for the neutron, where $\mu_N$ is the nuclear magneton
and $g_p=5.58$ and $g_n=-3.82$ are the Land\'{e} g-factors for the
proton and neutron, respectively.  The energy $|\kappa_n + \kappa_p| B
\simeq 1.67 \times 10^{-5} B^*$ MeV measures the changes in the beta
equilibrium condition and to the baryon Fermi energies.  Since the
Fermi energies range from a few MeV to tens of MeV for the densities
of interest, it is clear that contributions
from the anomalous magnetic moments also become significant for $B^* >
10^5$.  We demonstrate that for such fields, complete spin
polarization of the neutrons occurs, which results in an overall
stiffening of the EOS that overwhelms the softening induced by Landau
quantization.

In magnetized matter, the stress energy tensor contains terms
proportional to $HB$, where $H = B+4\pi{\cal M}$ and ${\cal M}$ is the
magnetization (Landau, Lifshitz \& Pitaevski\u{i} 1984).  Thus,
extra terms, in addition to the usual ones proportional to $B^2$,
are introduced into the structure equations (Cardall et al. 1999).
The magnetization in a
single component electron gas has been studied extensively (Blandford
\& Hernquist 1982) for neutron star crust matter. We generalize this
formulation to the case of interacting multicomponent matter with and
without the effects of the anomalous magnetic moments.  We find that
deviations of $H$ from $B$ occur for field strengths $B^* \gtrsim
10^5$.

Although the effects of magnetic fields on the EOS at low densities,
relevant for neutron star crusts, has been extensively studied
(see for example, Canuto \& Ventura 1977; Fushiki, Gudmundsson \& Pethick
1989; Fushiki et al. 1992; Abrahams \& Shapiro 1991;  Lai \& Shapiro 1991;
R\"{o}gnvaldsson et al. 1993, Thorlofsson et al. 1998),
only a handful of previous works have considered the
effects of very large magnetic fields on the EOS of dense neutron star
matter (Chakrabarty 1996; Chakrabarty, Bandyopadhyay, \& Pal
1997, Yuan \& Zhang 1999).  
Lai and Shapiro (1991) considered non-interacting, charge
neutral, beta-equilibrated matter at subsaturation densities, while
Chakrabarty and co-authors studied dense matter including interactions
using a field-theoretical description.  These authors found large
compositional changes in matter induced by ultra-strong magnetic
fields due to the quantization of orbital motion. Acting in concert
with the nuclear symmetry energy, Landau quantization substantially
increases the concentration of protons compared to the field-free
case, which in turn leads to a softening of the EOS. This lowers the
maximum mass relative to the field-free value.  In these works,
however, the electromagnetic field energy density and pressure, which
tend to stiffen the EOS, were not included. In addition, changes in the
general relativistic structure induced by the magnetic fields (studied
in detail by Bocquet et al. 1995 who, however, omitted the
compositional changes in the EOS due to Landau quantization) were also
ignored.  Thus, the combined effects of the magnetic fields on the EOS 
and on the general relativistic structure remain to be determined.

Compared to these earlier works, we make several improvements in the
calculation of the EOS. These improvements include (1) a study of a
larger class of field-theoretical models in order to extract the
generic trends induced by Landau quantization, (2) the development of
a covariant description for the inclusion of the anomalous magnetic
moments of the nucleons, and (3) a detailed study of magnetization of
interacting multicomponent matter with and without the inclusion of
the anomalous magnetic moments.  We also provide simple analytical
estimates of when each of these effects begin to significantly
influence the EOS.  Our future work will employ the EOSs developed in
this work to complete a fully self-consistent calculation of neutron
star structure including the combined effects of the direct effects of
magnetic fields on the EOS and general relativistic structure.

In \S2, we present the field-theoretical description of dense neutron
star matter including the effects of Landau quantization and the
nucleon anomalous magnetic moments. Section 3 contains a detailed
study of the effects of Landau quantization on the EOS for two classes
of Lagrangians.  In addition to providing contrasts with earlier work,
our results highlight the extent to which the underlying interactions
affect the basic findings.  This section also includes new theoretical
developments concerning the magnetization of interacting,
multicomponent matter. Section 4 is devoted to the effects of the
anomalous magnetic moments on the EOS. Here results for a charge
neutral neutron, proton, electron, and muon gas are compared with
those for interacting matter to asses the generic trends.  Our
conclusions and outlook, including the possible effects of additional
components such as hyperons, Bose condensates and quarks, are presented in
\S5.  The covariant description for the
inclusion of the anomalous magnetic moments of the nucleons is
presented in the Appendix, where explicit formulae for the nucleon
Dirac spinors and energy spectra are derived.  Except where necessary,
we use units wherin $\hbar$ and $c$ are set to unity.

\section{THEORETICAL FRAMEWORK}

For the description of the EOS of neutron-star matter,
we employ a field-theoretical approach in which the baryons (neutrons, $n$,
and protons, $p$)  interact
via the exchange of $\sigma~-\omega~-\rho$ mesons.  We study two classes of
models, which differ in their behavior at high density.
The Lagrangian densities associated with these two classes  are 
(Boguta \& Bodmer 1977, Zimanyi \& Moszkowski 1990)
\begin{eqnarray}
{\cal L}_{I} &=& {\cal L}_b -
\left( 1 - \frac{g_{\sigma_b} \sigma}{m_b}
\right) \overline{\Psi}_b m_b \Psi_b  + {\cal L}_m + {\cal L}_l \,,\nonumber\\
{\cal L}_{II} &=& \left( 1 + \frac{g_{\sigma_b} \sigma}{m_b} \right) {\cal L}_b
-  \overline{\Psi}_b m_b \Psi_b  + {\cal L}_m + {\cal L}_l \,.
\end{eqnarray}
The baryon ($b=n,p$), lepton ($l=e,\mu$), and meson
($\sigma,~\omega,~{\rm and}~\rho $) Lagrangians are given by
\begin{eqnarray}
{\cal L}_b &=& \overline{\Psi}_b \left( i \gamma_\mu \partial^\mu
+q_b \gamma_\mu A^\mu - g_{\omega_b} \gamma_\mu \omega^{\mu}
- g_{\rho_b} \tau_{3_b} \gamma_\mu \rho^{\mu}
- \kappa_b \sigma_{\mu \nu} F^{\mu \nu}
\right) \Psi_b \,,\nonumber \\
{\cal L}_l &=& \overline{\psi}_l \left( i \gamma_\mu \partial^\mu
+q_l \gamma_\mu A^\mu  \right) \psi_l \,,\nonumber \\
{\cal L}_m &=& \frac{1}{2} \partial_{\mu} \sigma \partial^{\mu} \sigma
- \frac{1}{2} m_{\sigma}^2 \sigma^2 - U(\sigma)
 + \frac{1}{2} m_{\omega}^2 \omega_{\mu} \omega^{\mu}
  - \frac{1}{4} \Omega^{\mu \nu} \Omega_{\mu \nu} \nonumber \\
 & & \mbox{} + \frac{1}{2} m_{\rho}^2 \rho_{\mu} \rho^{\mu}
- \frac{1}{4} P^{\mu \nu} P_{\mu \nu}  -\frac{1}{4} F^{\mu \nu} F_{\mu \nu} \,,
\end{eqnarray}
where $\Psi_b$ and $\psi_l$ are the baryon and lepton Dirac fields,
respectively.
The nucleon mass and the isospin projection are denoted by
$m_b$ and $\tau_{3_b}$, respectively.
The mesonic and electromagnetic  field strength tensors are given
by their usual expressions:  $\Omega_{\mu\nu} = \partial_\mu
\omega_\nu - \partial_\nu  \omega_\mu$,  $P_{\mu\nu} = \partial_\mu
\rho_\nu - \partial_\nu  \rho_\mu$,  and  $F_{\mu\nu} = \partial_\mu
A_\nu - \partial_\nu  A_\mu$.   The strong  interaction couplings are
denoted by $g$, the electromagnetic couplings by $q$, and the meson
masses by $m$ all with appropriate subscripts.
The anomalous magnetic moments are introduced via the coupling
of the baryons to the electromagnetic field tensor with
$\sigma_{\mu \nu} = \frac{i}{2} \left[\gamma_{\mu},\gamma_{\nu} \right]$
and strength $\kappa_b$.  We will contrast results for cases with
$\kappa_b=0$ and $\kappa_b$ taken to be their measured values.
The quantity $U(\sigma)$ denotes possible scalar self-interactions.
It is straightforward to include self interactions between both
 the vector $\omega$  and the iso-vector $\rho$ mesons (M\"uller \& Serot
1996). Although the electromagnetic field is included in
${\cal L}_I$ and ${\cal L}_{II}$, it assumed to be externally
generated (and thus has no associated field equation) and only
frozen-field configurations will be considered.

The thermodynamic quantities will be evaluated in the mean field
approximation, in which the mesonic fields are assumed to be constant. The
 field equations are
\begin{equation}
m_{\sigma}^2 \langle \sigma \rangle
+ \frac{\partial U(\sigma)}{\partial \sigma}
= \left\{ \begin{array}{cl}
\sum_b g_{\sigma_b} n^s_b & \mbox{for ${\cal L}_I$} \\
\sum_b g_{\sigma_b} \left(\frac{m_b^*}{m_b}\right)^2 n^s_b &
\mbox{for ${\cal L}_{II}$ \,,}
\end{array}
\right.
\label{feq:sig}
\end{equation}
\begin{equation}
m_{\omega}^2 \langle \omega^0 \rangle = \sum_b g_{\omega_b} n_b \,,
\label{feq:ome}
\end{equation}
\begin{equation}
m_{\rho}^2 \langle \rho^0 \rangle = \sum_b g_{\rho_b} \tau_{3_b} n_b \,,
\label{feq:rho}
\end{equation}
\begin{equation}
\left[ \vec{\alpha} \dotprod \left( \vec{p} - q_l \vec{A} \right)
+ \beta m_l \right]\psi_l = E \psi_l \,,
\label{feq:lep}
\end{equation}
\begin{equation}
\left[\vec{\alpha} \dotprod \left( \vec{p} - q_b \vec{A} \right)
+ \beta m_b^* \right] \Psi_b = (E - g_{\omega_b}
\omega^0 - g_{\rho_b} \tau_{3_b} \rho^0)\Psi_b \,,
\label{feq:bar}
\end{equation}
where the effective baryon masses are
\begin{equation}
\frac{m_b^*}{m_b} = \left\{ \begin{array}{cl}
\left( 1 - \frac{g_{\sigma_b} \sigma}{m_b} \right) & \mbox{for ${\cal L}_I$} \\
\left( 1 + \frac{g_{\sigma_b} \sigma}{m_b} \right)^{-1} &
\mbox{for ${\cal L}_{II}$ \,,}
\end{array} \right.
\end{equation}
and $n_b^s$ is the scalar number density.  The scalar
self-interaction is taken to be of the form 
(Boguta \& Bodmer 1977; Glendenning 1982, 1985) 
\begin{equation}
U(\sigma) = \frac{1}{3} b m_n (g_{\sigma_n} \sigma)^3
+    \frac{1}{4} c (g_{\sigma_n} \sigma)^4 \,,
\end{equation}
where the $m_n$ in the first term is included to make $b$ dimensionless.
In charge neutral, beta equilibrated matter, the conditions
\begin{equation}
\mu_n - \mu_p = \mu_e = \mu_{\mu} \,,
\label{chm}
\end{equation}
\begin{equation}
n_p = n_e + n_{\mu} \,,
\label{chg}
\end{equation}
also apply.
Given the nucleon-meson coupling constants and the coefficients in the scalar
self-interaction, equations (\ref{feq:sig}) through (\ref{chg}) may be
 solved self consistently for the
chemical potentials, $\mu_i$, and the field strengths,
$\sigma$, $\omega^0$, and $\rho^0$.

\section{EFFECTS OF LANDAU QUANTIZATION}

From equation (\ref{feq:lep}), the energy spectrum for the leptons is
(see, for example, Canuto \& Ventura 1977)
\begin{equation}
E_l = \sqrt{ k_z^2 + \widetilde{m}^{l~2}_{n,\sigma_z} } \,,
\end{equation}
where
\begin{equation}
\widetilde{m}_{n,\sigma_z}^{l~2} =  m_l^2 + 2 \left( n+\frac{1}{2}
-\frac{1}{2}\frac{q_l}{|q_l|}\sigma_z \right) |q_l| B \,.
\label{mtilde}
\end{equation}
Here, $n$ is the principal quantum number and $\sigma_z$ (not to be
confused with the scalar field $\sigma$) is the spin along the
magnetic field axis.  $k_z$ is the component of the momentum along the
magnetic field axis.  The quantity $\nu = n + 1/2 -
(1/2)(q_i/|q_i|)\sigma_z$ characterizes the so-called Landau level.
Equation (\ref{feq:bar}) gives the energy spectrum for the protons as
\begin{equation}
E_p = \sqrt{ k_z^2 + \widetilde{m}^{p~2}_{n,\sigma_z} } + g_{\omega_p} \omega^0
- g_{\rho_p} \frac{1}{2} \rho^0 \,,
\label{nmp:e}
\end{equation}
where  $\widetilde{m}^p_{n,\sigma_z}$ is obtained by replacing $m_l$ on the
right hand side of equation (\ref{mtilde}) by $m^*_p$.
The neutron energy spectrum is that of the free Dirac particle, but 
with shifts arising from the scalar, vector, and isovector 
interactions: 
\begin{equation}
E_n = \sqrt{ k^2 +m_n^{*~2} } + g_{\omega_n} \omega^0 + g_{\rho_n}\frac{1}{2}
\rho^0 \,.
\end{equation}

At zero
temperature and in the presence of a constant magnetic field $B$,
the number and energy densities of charged particles are given by
\begin{equation}
n_{i=l,p} = \frac{|q_i| B}{2 \pi^2} \sum_{\sigma_z} \sum_{n=0}^{n_{max}}
k_{f,n,\sigma_z}^i \,,
\label{nmax}
\end{equation}
\begin{equation}
\varepsilon_{i=l,p} = \frac{|q_i| B}{4 \pi^2} \sum_{\sigma_z} \sum_{n=0}^{n_{max}}
  \left[ E_f^i k_{f,n,\sigma_z}^i + \widetilde{m}_{n,\sigma_z}^{i~2} \ln
\left( \left|
\frac{E_f^i + k_{f,n,\sigma_z}^i}{\widetilde{m}^i_{n,\sigma_z}} \right|
\right) \right] \,.
\end{equation}
Above, $k_{f,n,\sigma_z}^i$ is the Fermi momentum for the level with the
principal quantum number $n$ and spin $\sigma_z$ and is given by
\begin{equation}
k_{f,n,\sigma_z}^{i~2} =  E_f^{i~2} - \widetilde{m}_{n,\sigma_z}^{i~2} \,.
\end{equation}
The summation in equation (\ref{nmax}) is terminated at
$n_{max}$, which is the integer preceeding the value of $n$ for which
$k_{f,n,\sigma_z}^{i~2}$ is negative.  The Fermi energies are fixed by the
chemical potentials
\begin{equation}
E_f^l = \mu_l \,,
\end{equation}
\begin{equation}
E_f^b = \mu_b - g_{\omega_b} \omega^0
- g_{\rho_b} \tau_{3_b} \rho^0 \,.
\end{equation}
For the protons, the scalar number density may be determined to be
(Chakrabarty 1996)
\begin{equation}
n_p^s = \frac{|q_p| B m^*_p}{2 \pi^2} \sum_{\sigma_z} \sum_{n=0}^{n_{max}}
\ln \left( \left|
\frac{E_f^p + k_{f,n,\sigma_z}^p}{\widetilde{m}_{n,\sigma_z}^p}
\right| \right) \,.
\end{equation}
The number, energy, and scalar number densities of the neutrons are
unchanged in form from the field-free case
\begin{equation}
n_n = \frac{k_f^{n~3}}{3 \pi^2} \,,
\end{equation}
\begin{equation}
n^s_n = \frac{ m_n^*}{2 \pi^2} \left[
E_f^n k_f^n - m_n^{*~2} \ln \left( \left|
\frac{E_f^n + k_f^n}{m_n^*} \right| \right) \right] \,,
\end{equation}
\begin{equation}
\varepsilon_n = \frac{1}{8 \pi^2} \left[ 2 E_f^{n~3} k_f^n - m_n^{*~2}
E_f^n k_f^n -m_n^{*~4} \ln \left( \left| \frac{E_f^n + k_f^n}{m_n^*}
\right| \right) \right] \,.
\end{equation}
The total energy density of the system is
\begin{eqnarray}
\varepsilon & = & \frac{1}{2} m_{\omega}^2 \omega_0^2
+ \frac{1}{2} m_{\rho}^2 \rho_0^2 + \frac{1}{2} m_{\sigma}^2 \sigma^2
+ U(\sigma) \nonumber \\
& & \mbox{} + \sum_b \varepsilon_b + \sum_l \varepsilon_l + \frac{B^2}{8 \pi^2} \,,
\end{eqnarray}
where the last term is the contribution from the electromagnetic
field.  Use of equations (10) and (11), which are satisfied in charge
neutral beta-equilibrated matter, in the general expression for the
pressure,  $P= \sum_i \mu_i n_i - \varepsilon$ ($i= n,p,e,~{\rm and}~
\mu$), allows the pressure to be written  only in terms of the neutron
chemical potential through the relation  $P=\mu_nn_b-\varepsilon$.  In
fact, utilizing the appropriate relations satisfied by the various chemical 
potentials and the number densities involved in the charge neutrality 
condition, it is easily verified that  
this relation is satisfied even in the presence of additional
components such as strangeness-bearing hyperons, Bose condensates 
(pion or kaon), and quarks,  which may
likely exist in dense neutron-star matter.

\subsection{Magnetization}

The magnetic field strength, $H$, is related to the energy
density by (Landau, Lifshitz \& Pitaevski\u{i} 1984)
\begin{equation}
H = 4 \pi \left( \frac{\partial \varepsilon}{\partial B} \right)_{n_b}
= B + 4 \pi {\cal M} \,,
\end{equation}
where ${\cal M}$ is the magnetization.
This is equivalent to the set of equations
\begin{eqnarray}
\frac {d n_b}{d B} &=& 0 \,, \nonumber
\\
{\cal M} &=&
\frac{\partial \varepsilon_l}{\partial B}
+
\frac{\partial \varepsilon_l}{\partial E_f^l} \frac{d E_f^l}{d B}
+
\frac{\partial \varepsilon_p}{\partial B}
+
\frac{\partial \varepsilon_p}{\partial E_f^p} \frac{d E_f^p}{d B}
+
\frac{\partial \varepsilon_n}{\partial B}
+
\frac{\partial \varepsilon_n}{\partial E_f^n} \frac{d E_f^n}{d B}
\nonumber\\
 & & \mbox{} +
m_{\rho}^2 \rho^0 \frac{d \rho^0}{d B}
+
m_{\omega}^2 \omega^0 \frac{d \omega^0}{d B}
+
\frac{\partial \varepsilon}{\partial \sigma} \frac{d \sigma}{d B} \,.
\end{eqnarray}
The first of these gives
\begin{equation}
\frac{\partial n_n}{\partial E_f^n} \frac{d E_f^n}{d B} =
-
\frac{\partial n_n}{\partial B}
-
\frac{\partial n_p}{\partial B}
-
\frac{\partial n_p}{\partial E_f^p} \frac{d E_f^p}{d B} \,.
\end{equation}
Using the conditions of charge neutrality and
chemical equilibrium, one has
\begin{equation}
\frac{\partial n_l}{\partial E_f^l} \frac{d E_f^l}{d B} =
-
\frac{\partial n_l}{\partial B}
+
\frac{\partial n_p}{\partial B}
+
\frac{\partial n_p}{\partial E_f^p} \frac{d E_f^p}{d B} \,.
\end{equation}
From the field equations and the definition of the scalar density,
\begin{eqnarray}
\frac{d \omega^0}{d B} &=& \frac{g_{\omega}}{m_{\omega}^2}
\frac{d n_b}{d B} = 0 \,,\nonumber
\\
\frac{d \rho^0}{d B} &=& \frac{g_{\rho}}{m_{\rho}^2} \frac{1}{2}
\frac{d}{d B} \left( n_n - n_p \right) =
-\frac{g_{\rho}}{m_{\rho}^2} \frac{d n_p}{d B} \,,
\nonumber \\
\frac{\partial \varepsilon}{\partial \sigma} &=& 0 \,.
\end{eqnarray}
Note also that
\begin{equation}
\frac{\partial \varepsilon_i}{\partial E_f^i} = E_f^i
\frac{\partial n_i}{\partial E_f^i} \,.
\end{equation}
Utilizing these results, equation (26) becomes
\begin{equation}
{\cal M} = T_{I}
+
T_{II} \frac{\partial n_p}{\partial E_f^p} \frac{d E_f^p}{d B} \,,
\end{equation}
where
\begin{eqnarray}
T_{I} &=&
\frac{\partial \varepsilon_l}{\partial B}
-
E_f^l \frac{\partial n_l}{\partial B}
+
\frac{\partial \varepsilon_n}{\partial B}
-
E_f^n \frac{\partial n_n}{\partial B}
+
\frac{\partial \varepsilon_p}{\partial B} \nonumber
\\
\mbox{} & &+
\left( E_f^l - E_f^n - g_{\rho} \rho^0 \right) \frac{\partial n_p}{\partial B}
\,, \nonumber \\
T_{II} &=&
E_f^l + E_f^p - E_f^n - g_{\rho} \rho^0 \,.
\end{eqnarray}
Note that chemical equilibrium ensures that $T_{II}=0$
whence the magnetization takes the general form
\begin{equation}
{\cal M} =  \sum_{i=e,\mu,p,n} \left( \frac {\partial \varepsilon_i}{\partial B}
- E_f^i \frac{\partial n_i}{\partial B} \right) \,.
\label{mag}
\end{equation}
In the case under current consideration, inserting the explicit forms
of the energy density and number density yields the result
\begin{equation}
{\cal M} =  \sum_{i=e,\mu,p} \left[
\frac{\varepsilon_i - E_f^i n_i}{B}
+
\frac{B}{2 \pi^2} \sum_{\sigma_z} \sum_{n=0}^{n_{max}}
\left( n + \frac{1}{2} - \frac{1}{2} \sigma_z  \right) \ln \left( \left|
 \frac{E_f^i + k_{f,n,\sigma_z}^i}{ \widetilde{m}_{n,\sigma_z}^i }
\right| \right) \right]\,.
\end{equation}
This result generalizes the result of Blandford \& Hernquist (1982) for
an electron gas to the case of a multi-component system including
interacting nucleons.  That the functional form of ${\cal M}$ for
interacting nucleons is the same as that for non-interacting particles
stems from the fact that, in the mean field approximation, the field
equations for the nucleons reduces to
the Dirac equation for a free
particle, but with an effective mass $m^*$.

\subsection{Results}

In Table 1, we list the various nucleon-meson and meson
self-interaction couplings for the two classes of models chosen for
this study.  In each case, the couplings were chosen to reproduce
commonly accepted values of the equilibrium nuclear matter properties:
the binding energy per particle $B/A$, the saturation density $n_s$,
the Dirac effective mass $m_n^*/m_n$, the compression modulus $K_0$,
and the symmetry energy $a_{\mbox{\tiny sym}}$.  The high-density
behavior of the EOS is sensitive to the strength of the meson
couplings employed and the models chosen encompass a fairly wide range
of variation.  The HS81 model, which has a rather high compression
modulus, allows us to contrast our results with those of Chakrabarty
(1996) who also used HS81 in the case when Landau quantization is
considered, and to assess the effects of the inclusion of magnetic
moments.  Models HS81 and GM1--GM3 employ linear scalar couplings
(${\cal L}_I$), while the ZM model employs a nonlinear scalar coupling
(${\cal L}_{II}$), which is reflected in the high density behaviors of
$m_n^*/m_n$.  Thus, comparison of the HS81, GM1--GM3 and ZM models
allows us to contrast the effects of the underlying EOS.

In Figure \ref{fig1}, we show results of some physical quantitites of
interest for our baseline case, model GM3.  At supernuclear densities
and in the absence of a magnetic field, the matter pressure is
dominated by the baryons principally due to the repulsive nature of
the strong interactions.  Even up to the central density in a neutron
star, the proton fraction remains sufficiently small that the neutrons
dominate the total pressure.

The magnitude of the magnetic field $B$ required to induce significant
changes in the EOS may be estimated in a straightforward manner.  In
the presence of a magnetic field, the contributions from the protons
become significant when only one Landau level is occupied, i.e., when
the protons are completely spin polarized.  This happens when $q_p
B/\hbar c > (2 \pi^4 n_p^2 )^{1/3}$.  Therefore, we arrive at
the estimate $B^*>\lbar^2 (2 \pi^4 Y_p^2 n_b^2)^{1/3}$ for
quantum effects to dominate.  The proton fraction $Y_p=n_p/n_b$, which
depends upon both the density and the magnetic field, typically lies
in the range 0.1--0.7.  As a result the term in parentheses is of
order $1~\mbox{fm}^{-2}$ for densities $n_b > 0.1 n_s$, and $\lbar^2
\simeq 1.5\times10^5 \; \mbox{fm}^{2}$.  Thus, the magnetic field
necessary to introduce significant contributions from the protons is
of order $B^*\simeq 10^5$, which is well below the proton critical
field $B_c^p = (m_p/m_e)^2B_c^e=1.49\times 10^{20}$ Gauss (or
$B^*=3\times10^6$) for which protons begin to become relativistic.

The results in Figure \ref{fig1} were obtained by
accounting for all of the  allowed
Landau levels.   Indeed, we notice that the matter pressure  $P_m$,
the effective mass $m_n^*$, and the  concentrations $Y_i=n_i/n_b$
begin to differ significantly  from their field-free values  only for
$B^* \simge  10^5$.  The results in the right panels, shown as a
function  of $B^*$ for four values of $u=n_b/n_s$, show that the
density dependence of this threshold value is also qualitatively correct.

The upper left panel shows that  there is a substantial decrease
in the pressure associated with increasing magnetic fields for $B^* >
10^5$.   This is also evident from the inset, which
clearly shows extensive softening of the EOS.  The onset of
changes in the pressure  as a function of the magnetic field may be
more clearly seen in the upper right panel, in which  results for
representative densities are shown.

The neutron effective mass $m_n^*$ is shown in the lower left panel,
and demonstrates the extent to which the scalar field $\sigma$ is
influenced by the presence of magnetic fields.  Note that $m_n^*$
also enters in the calculation of all thermodynamic quantities.
Again, it is clear that effects due to magnetic fields do not become
significant until $B^* > 10^5$.

The lower right panel shows that the composition of neutron-star
matter changes significantly at high magnetic fields.  The striking
feature is the large increase in the proton fraction for $B^* \simge
10^5$.  This has two significant effects upon the EOS.  First, the
protons, which are spin polarized, begin to dominate the contributions
to thermodynamics arising from the baryons.  This leads to a
substantial softening of the EOS (see upper left panel).  The second
effect stems from the requirement of charge neutrality.  Because the
leptons provide the only source of negative charge, the lepton
fraction rises commensurately with the proton fraction.  As a result,
the lepton contributions to the pressure and energy density are
somewhat increased relative to the field-free case.  However, the
contributions from the baryons remain dominant.

It is important to note that in order to obtain the total energy
density and presssure relevant for neutron star structure,
contributions from the electromagnetic field $\varepsilon_f = P_f =
B^2/(8\pi) = 4.814 \times 10^{-8} B^{*~2}~{\rm MeV~ fm}^{-3}$ must be
added to the matter energy density $\varepsilon_m$ and pressure $P_m$.
This has not always been done in the literature.  For $B^*>10^5$,
 the field contributions can dominate the
matter pressure,  for
the densities of interest, as shown in the upper right panel of Figure 1.

Figure \ref{fig2} shows the dependance of $H/B$ on the field strength
$B^*$ and the density $u$ for the baseline model GM3.  The so-called
de Haas-van Alphen oscillations are evident and highlight
the multi-component nature of the system.  The origin of the
increasing complexity in the oscillations may be understood by first
inspecting the oscillation period when only a single charged species
is present.  When the quantity $(E_f^2-m^2)/(2qB)$ successively
approaches integer values, successive Landau orbits begin to get
populated resulting in an oscillatory structure in $H/B$. The width
of these oscillations may be estimated by considering the change in magnetic
field $\Delta B$ required to increase $n_{max}$ by 1.   It is found
to be dependent upon both the strength of the magnetic
field and the Fermi momenta, and is given by
\begin{equation}
\Delta B = B_+-B_-=\frac {2 qB_+^2}{k_f^2 +2 qB_+} \,,
\end{equation}
where $B_-$ and $B_+$ denote the fields at the beginning and end of an
oscillation.  In the low field limit
\begin{equation}
\Delta B \rightarrow 2q\left(\frac{B_+}{k_f}\right)^2 \,,
\end{equation}
and the period goes to zero.  At subnuclear densities, where muons are
generally absent, charge neutrality forces the Fermi momenta of
protons and electrons to be equal and only a single oscillation
period exists.  However, as the density increases above nuclear densities,
the appearance of muons introduces further structure in the
oscillations as a result of the superposition arising from
each of the three charged species present.  Furthermore, with
increasing density the Fermi momenta of all particle species increase,
which decreases the oscillation periods.  The insets in each of the
panels clearly show these features.

At large enough magnetic fields, only one Landau level is occupied,
and the value of $H/B$ saturates.  Beyond this point, the fraction of
$H$ that the magnetization comprises becomes increasingly small.  This
is demonstrated in the lower right panel of Figure \ref{fig2}, in
which the ratio $H/B$ approaches unity for both $B^* = 0$ and
$B^*\simeq3.3\times10^6 \simeq B^p_c$.  However, for $B^*\simeq10^5$
there is a noticeable deviation of $H/B$ from unity.  Nevertheless, in
all cases considered, $|4\pi{\cal M}/B|$ does not exceed 4\%, which does
not represent a significant deviation from the case in which the
magnetization is neglected.

In Figure \ref{fig3}, we compare results of the matter pressure and
effective mass for the models HS81 (employed earlier by Chakrabarty
(1996)), GM1, GM2, and ZM with a view towards extracting generic
trends induced by strong magnetic fields.  Although quantitative
differences persist between the models, the qualitative trends of the effects
of the field on
these EOSs are shared with those of model GM3.   Remaining differences are
principally due to variations in the underlying stiffnesses,
effective masses, and symmetry energies of the individual models.

\section{EFFECTS INCLUDING ANOMALOUS MAGNETIC MOMENTS}

We turn now to the inclusion of  the anomalous magnetic moments of the
nucleons.   Johnson \& Lippman (1950) first considered  the
inclusion of anomalous magnetic moments in the Dirac equation, but
their formulation was noncovariant. Here, we employ   the covariant
form,  suggested by Bjorken \& Drell (1964), using ${\cal L}_I$ and
${\cal L}_{II}$  to evaluate the effects of magnetic fields.  With the
inclusion of the anomalous magnetic moments,  the baryon field
equations become
\begin{equation}
\left[\vec{\alpha} \dotprod \left( \vec{p} - q_b \vec{A} \right)
+ \beta m_b^* \right] \Psi_b = \left(E - g_{\omega_b}
\omega^0 - g_{\rho_b} \tau_{3_b} \rho^0  +
\kappa_b
\frac{i}{2} \gamma_0 \left[\gamma_{\mu},\gamma_{\nu} \right] F^{\mu \nu}
\right)\Psi_b \,.
\label{kapeq}
\end{equation}
The derivation of the Dirac spinors is presented in Appendix A.

The energy spectrum for the protons is given by
\begin{eqnarray}
E_{p,n,s} &=& \sqrt{ k_z^2 + \left( \sqrt{m_p^{*~2} +
2 \left(n+\frac{1}{2} + s\frac{1}{2} \right) q_p B}
+ s \kappa_p B \right)^2 }
 + g_{\omega_p} \omega^0 - \frac{1}{2} g_{\rho_p} \rho^0
\label{ep_kap}
\end{eqnarray}
which may be compared with the result
\begin{eqnarray}
E_{p,n,s}^{JL} &=& \sqrt{2 \left(n+\frac{1}{2} + s\frac{1}{2} \right) q_p B
 + \left( \sqrt{m_p^{*~2} + k_z^2 }
+ s \kappa_p B \right)^2 }
 + g_{\omega_p} \omega^0 - \frac{1}{2} g_{\rho_p} \rho^0
\end{eqnarray}
obtained by using the
Johnson \& Lippman form
$ \kappa_b (i/2) \left[\gamma_{\mu},\gamma_{\nu} \right]
F^{\mu \nu} $
in equation (\ref{kapeq})
for the inclusion of the magnetic moment.  In both cases $n$ and $s$ are the
principle quantum number and ``spin'' quantum number respectively.  As will
be shown in the appendix, unlike the $\kappa_b=0$ case, the ``big'' components
of the Dirac spinor are no longer eigenstates of the spin operator along the
magnetic field ($\sigma_z$).  However, as
$\kappa_p$ tends toward zero, it is clear that the proton energy spectum
reduces to the expression given in equation (\ref{nmp:e}) and $s$
corresponds to the $\sigma_z$ eigenvalue.
In the nonrelativistic limit, $k_z^2 \ll m^2$ and
$2\nu q_pB \ll m^2$ and for $(\kappa_p B)^2 \ll m^2$,
both of the above results reduce to
$ E_{p,n,s} \simeq m_p^* + k_z^2/2 m_p^*
+ \left(n + 1/2 + s/2 \right) (q_p B)/m_p^* + s \kappa_p B$, the
standard nonrelativistic expression.

The evaluation of the thermodynamic quantities proceeds along the
lines already presented in \S2, but with a new definition of
the Fermi momentum
to account for the presence of magnetic moments, which cause an
asymmetry in the phase space in addition to that caused by the
charged particle interactions with $B$. With the energy spectrum in
equation (\ref{ep_kap}),
\begin{equation}
k_{f,n,s}^p = \sqrt{ E_f^{p~2} - \left( \sqrt{m_p^{*~2} + 2 \left(
n+\frac{1}{2} + \frac{1}{2} s \right) q_p B} + s \kappa_p B \right)^2} \,.
\end{equation}
Setting
\begin{equation}
\overline{m} = \sqrt{m_p^{*~2} + 2 \left( n+\frac{1}{2} +
\frac{1}{2} s \right)
q_p B} + s \kappa_p B\,,
\end{equation}
and substituting $k_{f,n,s}$ and $\overline{m}$ into the formulas for the number and
energy density for the leptons (see \S~2),
one finds the analogous quantities for the protons:
\begin{eqnarray}
n_p &=& \frac{|q_p| B}{2 \pi^2} \sum_n \sum_s k_{f,n,s}^p \,,\\
\varepsilon_p &=& \frac{|q_p| B}{2 \pi^2} \sum_n \sum_s E_f^p k_{f,n,s}^p +
\overline{m}^2 \ln \left( \left|
\frac{E_f^p + k_{f,n,s}^p}{\overline{m}} \right| \right) \,.
\end{eqnarray}
The scalar density may be defined in terms of the energy spectrum by
\begin{equation}
n^s = \int d^3 k  \frac{\partial  E }{\partial m_p^*}\,.
\end{equation}
Utilizing equation (\ref{ep_kap}) results in
\begin{eqnarray}
n^s_p &=& \int d^3 k  \frac{ \overline{m}}{\overline{m} - s \kappa_p B} \frac{m_p^*}{E}  \nonumber \\
&=& \frac{|q_p| B}{2 \pi^2} \sum_n \sum_s m_p^*
\frac{ \overline{m}}{\overline{m} - s \kappa_p B}
\ln \left( \left| \frac{E_f^p +
k_{f,n,s}^p}{ \overline{m}} \right| \right) \,.
\end{eqnarray}
The appearance of the factor ${ \overline{m}}
/({\overline{m} - s \kappa_p B})$ may be understood by inspecting the zeroth component of the current four-vector
\begin{equation}
\Psi^{p \dagger} \Psi^p = j^0_p= \gamma ~j^0_p|_{k=0} =
\gamma \overline{\Psi^p} \Psi^p
\end{equation}
with
\begin{equation}
\gamma = \frac{E}{E|_{k=0}} = \frac{E}{m_p^* + s \kappa_p B'}\,,
\end{equation}
where $B'$ is taken in the rest frame of the particle.
Hence, the scalar density becomes
\begin{equation}
n^s_p = \langle\overline{\Psi^p} \Psi^p\rangle
=\left\langle\frac{m_p^* + s \kappa_p B'}{E} \Psi^{p \dagger}
\Psi^p\right\rangle
=\int d^3 k \frac{\overline{m}}{\overline{m} - s \kappa_p B} \frac{m_p^*}{E}\,.
\end{equation}

In a similar way, the energy spectrum of the neutrons is given by
\begin{equation}
E_{n,s} = \sqrt{ k_z^2 + \left( \sqrt{m_n^{*~2} + k_x^2 + k_y^2} +
s \kappa_n B \right)^2}  + g_{\omega_n} \omega^0 +
\frac{1}{2} g_{\rho_n} \rho^0\,.
\label{en_kap}
\end{equation}
Including the magnetic moment for the neutron, the
integral over phase space for any
thermodynamical quantity $Q$ may be easily evaluated by noting that
at zero temperature, it is simply the integral
over all momenta within the Fermi surface defined by
\begin{equation}
E_f^n = E_{n,s} ( k_x,k_y,k_z) \,.
\end{equation}
The integral may be written in terms of parallel and perpendicular components,
\begin{equation}
\langle Q \rangle  = \sum_s
\frac{1}{2 \pi^2} \int_0^b k_{\perp} d k_{\perp} \int_0^a d k_{\parallel}~Q\,,
\end{equation}
where $a$ and $b$ are determined by the Fermi surface to be
\begin{eqnarray}
a &=& \sqrt{ E_f^{n~2} - \left(
\sqrt{ k_{\perp}^2 + m_n^{*~2}} + s \kappa_n B \right)^2} \,,\\
b &=& \sqrt{ (E_f^n - s \kappa_n B )^2 - m_n^{*~2} }\,.
\end{eqnarray}
With the substitution
\begin{eqnarray}
x = \sqrt{ k_{\perp}^2 + m_n^{*~ 2}} + s \kappa_n B \,,
\end{eqnarray}
the integral is transformed into
\begin{equation}
\langle Q \rangle = \sum_s \left(
\int_{\overline{m}}^{E_f^n} x dx \int_{0}^{\sqrt{E_f^{n~2} - x^2}}
d k_{\parallel}~Q \right) -
s \kappa_n B \left( \int_{\overline{m}}^{E_f^n} dx
\int_{0}^{\sqrt{E_f^{n~2} - x^2}} d k_{\parallel} ~Q \right) \,,
\end{equation}
where
\begin{equation}
\overline{m} = m_n^* + s \kappa_n B\,.
\end{equation}
Note that the first term is precisely the same as for the $\kappa_n = 0$ case,
but with a shifted mass.
This form for the integral over phase space is particularly useful
for calculating the number and energy densities.  Defining $k_{f,s}$ by
\begin{equation}
k_{f,s} = \sqrt{ E_f^{n~2} - \overline{m}^2} \,
\end{equation}
the number and energy densities take the form
\begin{eqnarray}
n_n &=& \frac{1}{2 \pi^2}  \sum_s \frac{1}{3} k_{f,s}^3 +
\frac{1}{2} s \kappa_n B \left[ \overline{m} k_{f,s} + E_f^{n~2} \left(
\arcsin \frac{\overline{m}}{E_f^n} - \frac{\pi}{2} \right) \right]\,, \\
\label{num_kap}
\varepsilon_n &=& \frac{1}{4 \pi^2} \sum_s \frac{1}{2} E_f^{n~3} k_{f,s}
 + \frac{2}{3} s \kappa_n B E_f^{n~3} \left(
\arcsin \frac{\overline{m}}{E_f^n} - \frac{\pi}{2} \right) \nonumber \\
&+& \left( \frac{1}{3} s \kappa_n B - \frac{1}{4} \overline{m} \right) \left[
\overline{m} k_{f,s} E_f^n + \overline{m}^3 \ln \left(\left| \frac{ E_f^n + k_{f,s}}
{\overline{m}}\right| \right) \right]\,.
\end{eqnarray}
The scalar number density reads
\begin{equation}
n^s_n = \int d^3 k \left( 1 + \frac{ s \kappa_n B}{\sqrt{ k_{\perp}^2 +
m_n^{*~2}}}
\right) \frac{m_n^*}{E} \,,
\end{equation}
which may be recast as
\begin{equation}
n^s_n = \sum_s
\int_{\overline{m}}^{E_f^n} x dx \int_{0}^{\sqrt{E_f^{n~2} - x^2}}
d k_{\parallel} \frac{m_n^*}{\sqrt{ k_{\parallel}^2 + x^2}}\,.
\end{equation}
Performing the integration gives
\begin{equation}
n^s_n = \frac{m_n^*}{4 \pi^2} \sum_s k_{f,s} E_f^n - \overline{m}^2
\ln \left( \left| \frac{ E_f^n + k_{f,s}} {\overline{m}}\right|
\right) \,.
\end{equation}
As in the case without magnetic moments, the pressure in beta
equilibrium is given by $P = \mu_n n_b - \varepsilon$.

\subsection{Magnetization}

Utilizing the expressions for the energy and number
densities derived above, the
magnetization including the effects of the anomalous magnetic moments may
be calculated using the general relation in equation (\ref{mag}).
For protons and neutrons, the results are given by
\begin{eqnarray}
{\cal M}_p &=&  \left\{ \frac{ \varepsilon_p - E_f^p n_p }{ B } +
\frac{1}{2 \pi^2} \sum_n \sum_s \overline{m} \ln \left( \left|
\frac{ E_f^p + k_{f,n,s}^p}{\overline{m}} \right| \right)
\left[ \frac{ \left( n + \frac{1}{2} + \frac{1}{2} s \right) }
{ \overline{m} - s \kappa_p B} + s \kappa_p B
\right] \right\}\,, \\
{\cal M}_n &=&  \frac{1}{2 \pi^2} \sum_s \kappa_n s \left\{ \left( \frac{1}{6}
\overline{m} -\frac{1}{2} s \kappa_n B \right) E_f^n k_{f,s} - \frac{1}{6} E_f^{n~3}
\left( \arcsin \frac{\overline{m}}{E_f^n} - \frac{\pi}{2} \right)
\right. \nonumber \\
&&\mbox{} + \left. \left( \frac{1}{2} s \kappa_n B - \frac{1}{3} \overline{m} \right)
\overline{m}^3 \ln \left( \left| \frac{ E_f^n + k_{f,s}}{\overline{m}} \right| \right)
\right\}\,.
\end{eqnarray}
The extent to which the anomalous magnetic moments alter the magnetization relative to
the case in which they are absent may be gauged by the magnitudes of
$H/B$ in single component systems.
For example, at fields below $2.2 \times 10^{19}$ Gauss,
$H/B$ is reduced by approximately 1\% in a proton gas and by
about  0.3\% in a neutron gas.

\subsection{Results for the $npe\mu$ Gas}

To assess the influence of the anomalous magnetic moments on the EOS,
it is instructive to consider a charge neutral $npe\mu$ gas in beta equilibrium.
In addition to providing contrasts with the case in which only the effects of Landau
quantization are considered (Lai \& Shapiro 1991),
it sets the stage for the effects to be expected for the case
in which baryonic interactions are included.

The magnitude of the magnetic field required to induce significant effects
on the EOS due to the inclusion of the magnetic moments may be inferred by
considering the field strength at which neutrons become completely polarized.
From equation  (\ref{num_kap}), it is clear that complete polarization occurs
when $|\kappa_n| B = k_{f,+1}^2/(4m_n) \simeq (6\pi^2n_n)^{\frac 23}/(4m_n)$.
At nuclear density, this leads to $B^* \cong 1.6 \times 10^5$.
Note that this is approximately where the effects due to Landau quantization
become large.  This implies that a complete description of neutron-star matter
in the presence of intense magnetic fields must necessarily include the
nucleon anomalous magnetic moments.

The equations governing the thermodynamics of the gas are simply the
non-interacting limits of equations (\ref{feq:sig}) through (\ref{chg}), and
equation (\ref{mag}) for the magnetization.
The results are presented in Figure \ref{fig4} in which  the darker (lighter)
shade curves show results with (without) the inclusion of
the anomalous magnetic moments.

The left panels, in which the matter pressure is shown  as functions
of $u$ and $\varepsilon_m$, clearly show that the EOS is stiffened
upon the inclusion of magnetic moments.  For example, in the extreme
case when the field strength
approaches  the proton critical field, the pressure is increased
by an order of  magnitude over the zero field case (and two orders of
magnitude over the case in which only the effects of Landau
quantization are considered).    The upper right  panel, in which  the
matter pressure is shown  as a function of $B^*$,
shows that  above $B^* = 10^5$ the effects  of the magnetic moments are
more significant  than those due to Landau quantization, and cannot be
ignored.

The lower right panel provides some insight into the origin of the
stiffening.  At field strengths of $B^*=10^5$,  the composition of
matter is dominated by neutrons,  the proton fraction being small,
about 0.1. Neutrons, however, are spin (up) polarized due to the
interaction of the magnetic moment with the magnetic field.
With increasing $B$, the fraction of neutrons that are polarized
increases leading to a corresponding increase in the degeneracy pressure.
Upon complete polarization, this increase is halted due to the absence of
neutrons needed to fill further spin up energy levels.
This is evident from the turnover in the matter pressure, occuring precisely
at the point when the neutrons become completely spin-polarized,
shown in the upper right panel.

\subsection{Results for Interacting Matter}

In this section, we include the effects of baryonic interactions,
Landau quantization, and anomalous magnetic moments.   In the absence
of magnetic fields, the dominant effect of interactions  between the
baryons is to substantially stiffen the EOS compared to the  case in
which interactions are omitted. This is chiefly due to the repulsive
nature of the baryonic interactions in beta stable matter.
Notwithstanding the fact that  the absolute magnitudes of the energy
density and pressure  are larger than the  case in which  the baryonic
interactions are omitted, magnetic fields have many of the  the
qualitative effects discussed in the previous section.

\begin{figure}
\begin{center}
\epsfxsize=6.in
\epsfysize=7.in
\epsffile{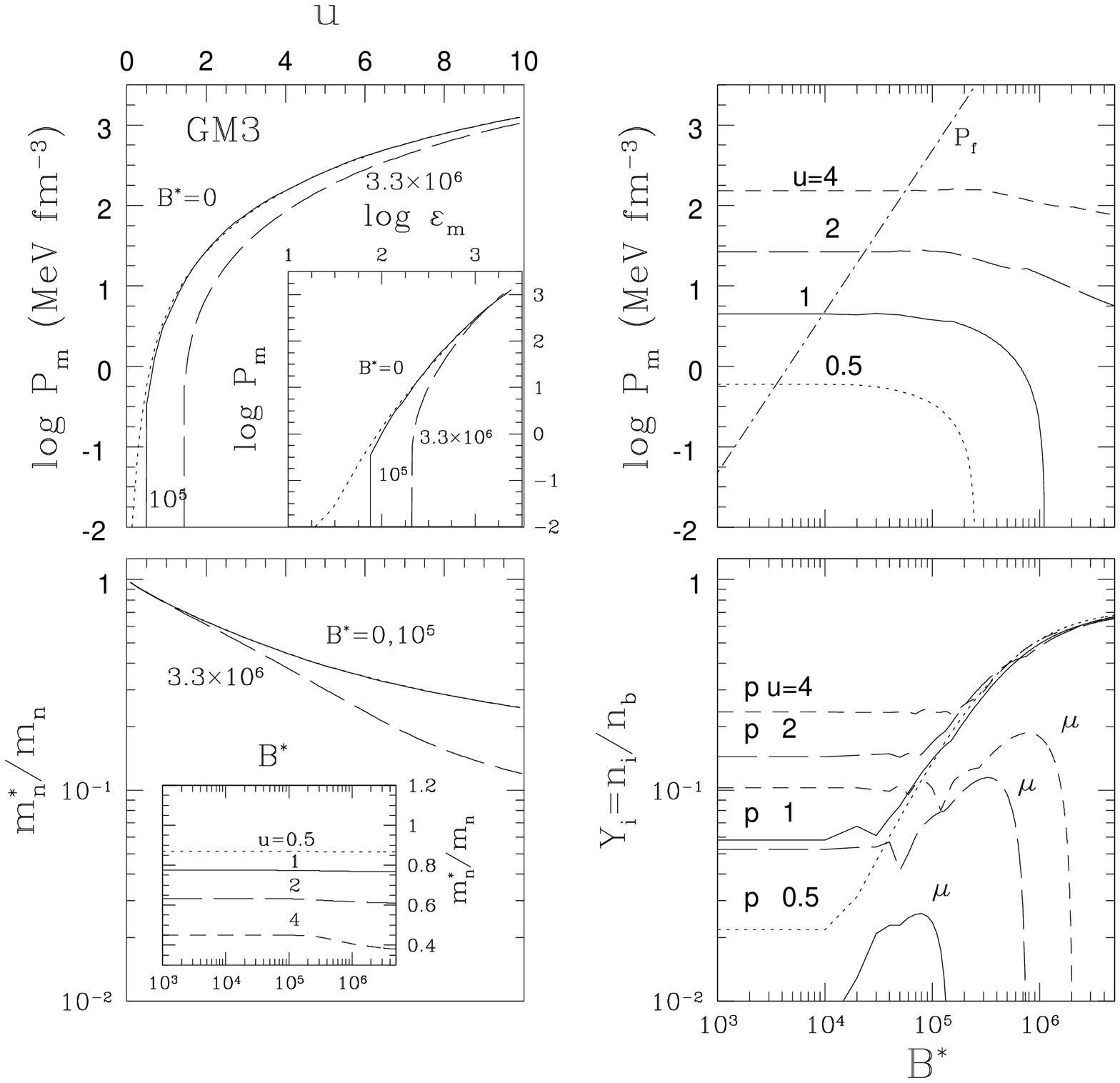}
\end{center}
\caption{}
{\label{fig1}}
\end{figure}

\begin{figure}
\begin{center}
\epsfxsize=6.in
\epsfysize=7.in
\epsffile{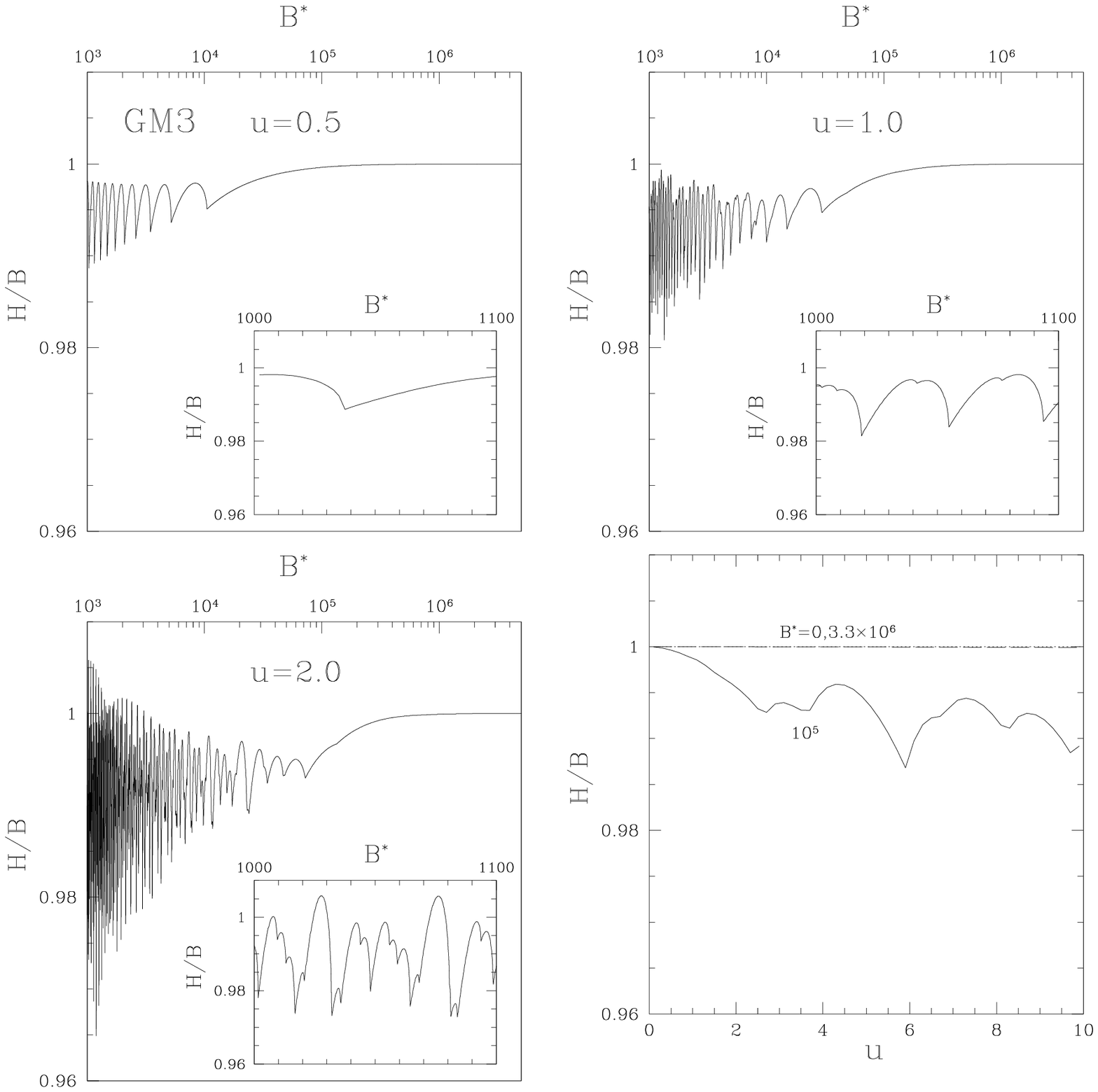}
\end{center}
\caption{}
{\label{fig2}}
\end{figure}

\begin{figure}
\begin{center}
\epsfxsize=6.in
\epsfysize=7.in
\epsffile{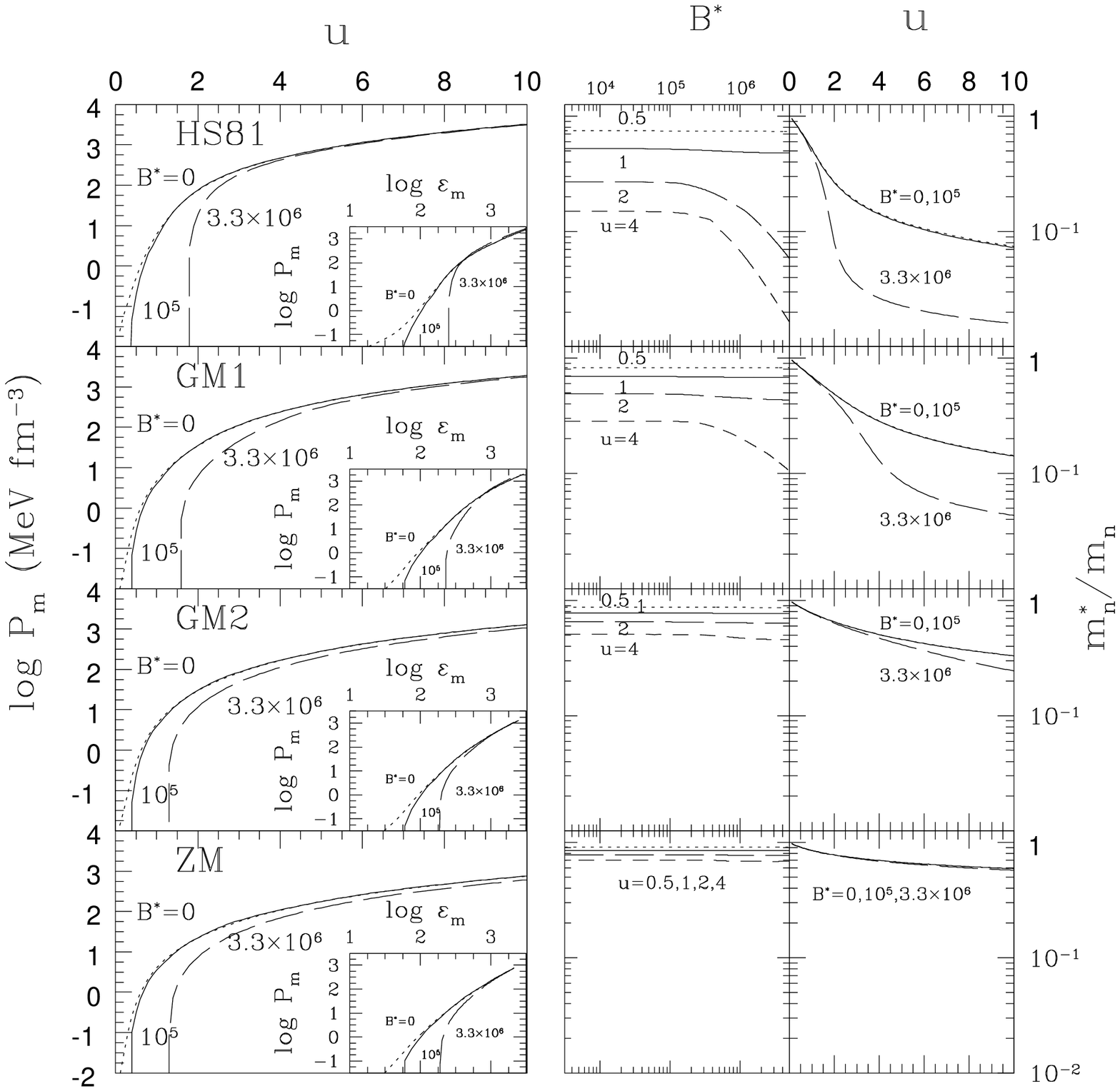}
\end{center}
\caption{}
{\label{fig3}}
\end{figure}

\begin{figure}
\begin{center}
\epsfxsize=6.in
\epsfysize=7.in
\epsffile{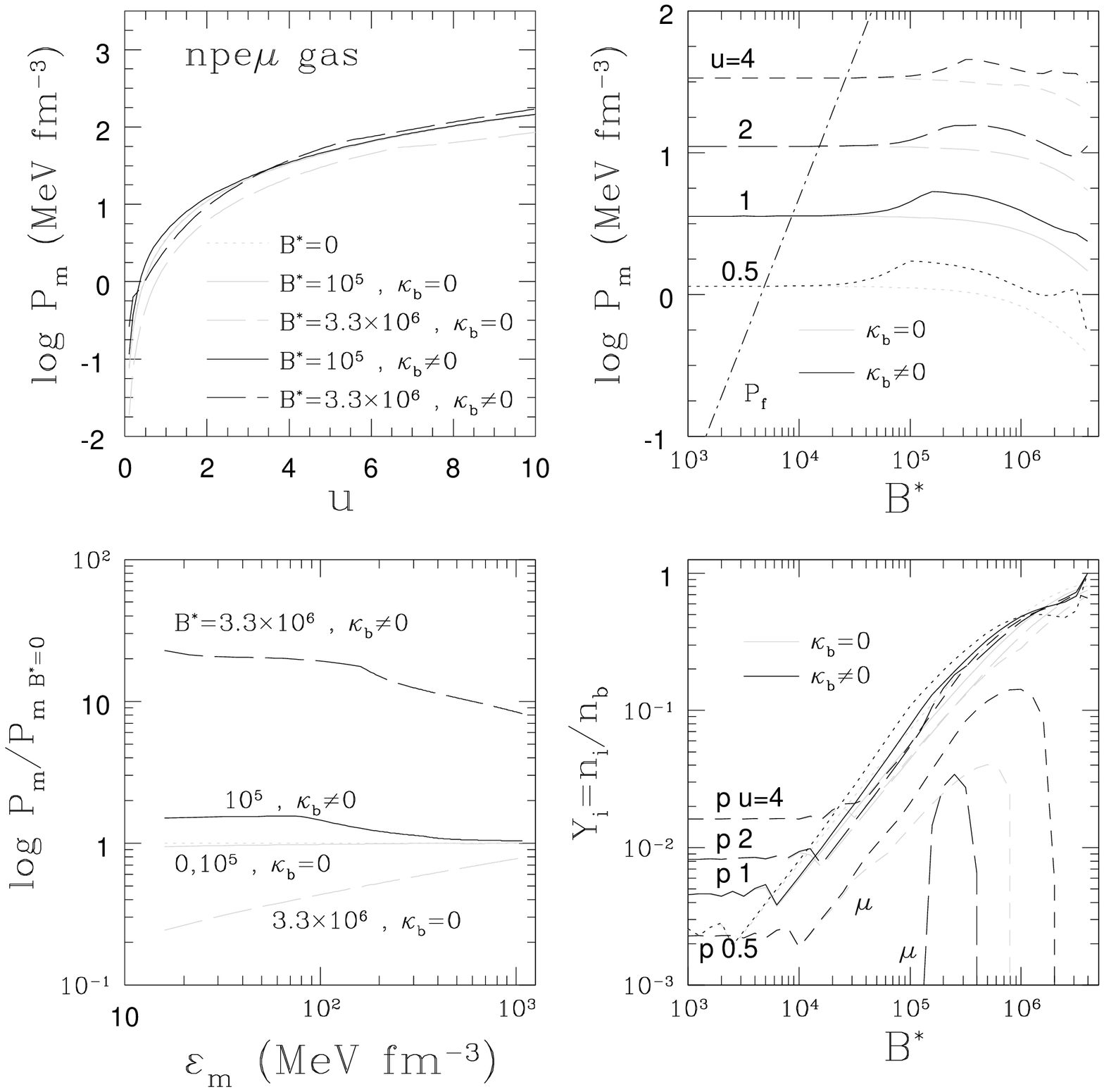}
\end{center}
\caption{}
{\label{fig4}}
\end{figure}

\begin{figure}
\begin{center}
\epsfxsize=6.in
\epsfysize=7.in
\epsffile{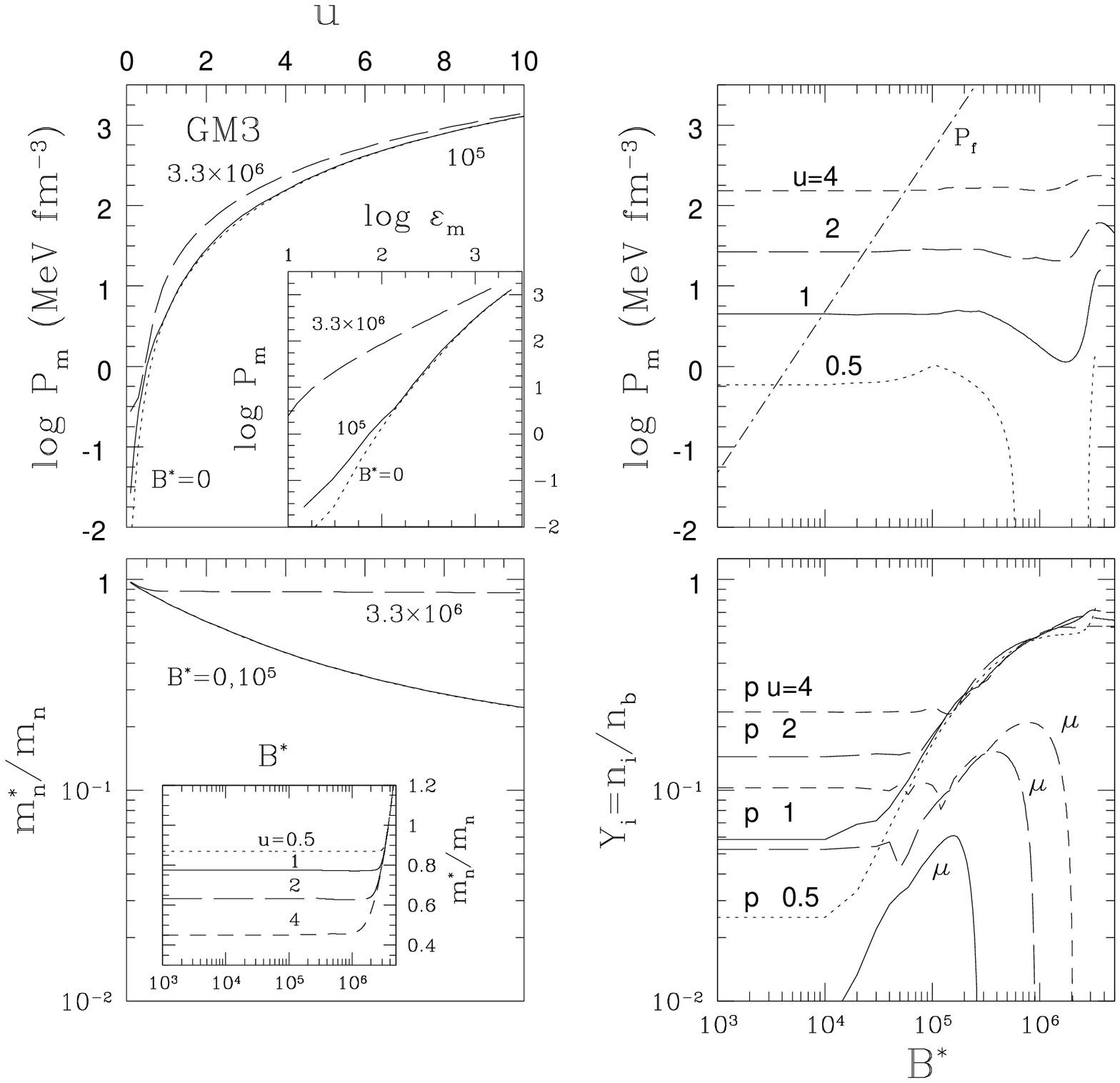}
\end{center}
\caption{}
{\label{fig5}}
\end{figure}

The results for
the baseline model GM3 are shown in Figure \ref{fig5}, which should be
compared with Figure \ref{fig1} to assess the role of magnetic
moments.  The upper left panel shows that the stiffening of the EOS
observed for the $npe\mu$ gas (for $B^* > 10^5$) is also present in the
case when interactions are included.  The effects of magnetic moments
are such that the softening caused by Landau quantiziation alone is
overwhelmed, leading to an overall stiffening of the EOS.  In fact, for
fields on the order of the critical proton field, the EOS approaches the
causal limit, $p_m=\varepsilon_m$.  As in Figure \ref{fig1}, the
matter pressure $P_m$, the effective mass $m_n^*$, and the
concentrations $Y_i=n_i/n_b$ begin to differ significantly from their
field-free values only for $B^* \simge 10^5$.

The neutron effective mass $m_n^*$ is shown in the lower left panel.
The behavior of $m_n^*$ with $B^*$ is opposite to that shown in Figure
\ref{fig1}.
The effects of magnetic moments cause $m_n^*$ to increase at a rate
approximately equal to $\kappa_p B/m_n$ and to become
independent of density for $B^* > 10^6$.  Note that this feature is also a
consequence of complete spin polarization.

The lower right panel shows the relative concentrations.
Comparing with Figure \ref{fig1}, it is evident that the composition of matter
is principally controlled by the effects of Landau quantization.
In contrast,  the
stiffening of the EOS is caused primarily by
terms that are explicitly dependent upon the
magnetic moments in the pressure and energy density.

\begin{figure}
\begin{center}
\epsfxsize=6.in
\epsfysize=7.in
\epsffile{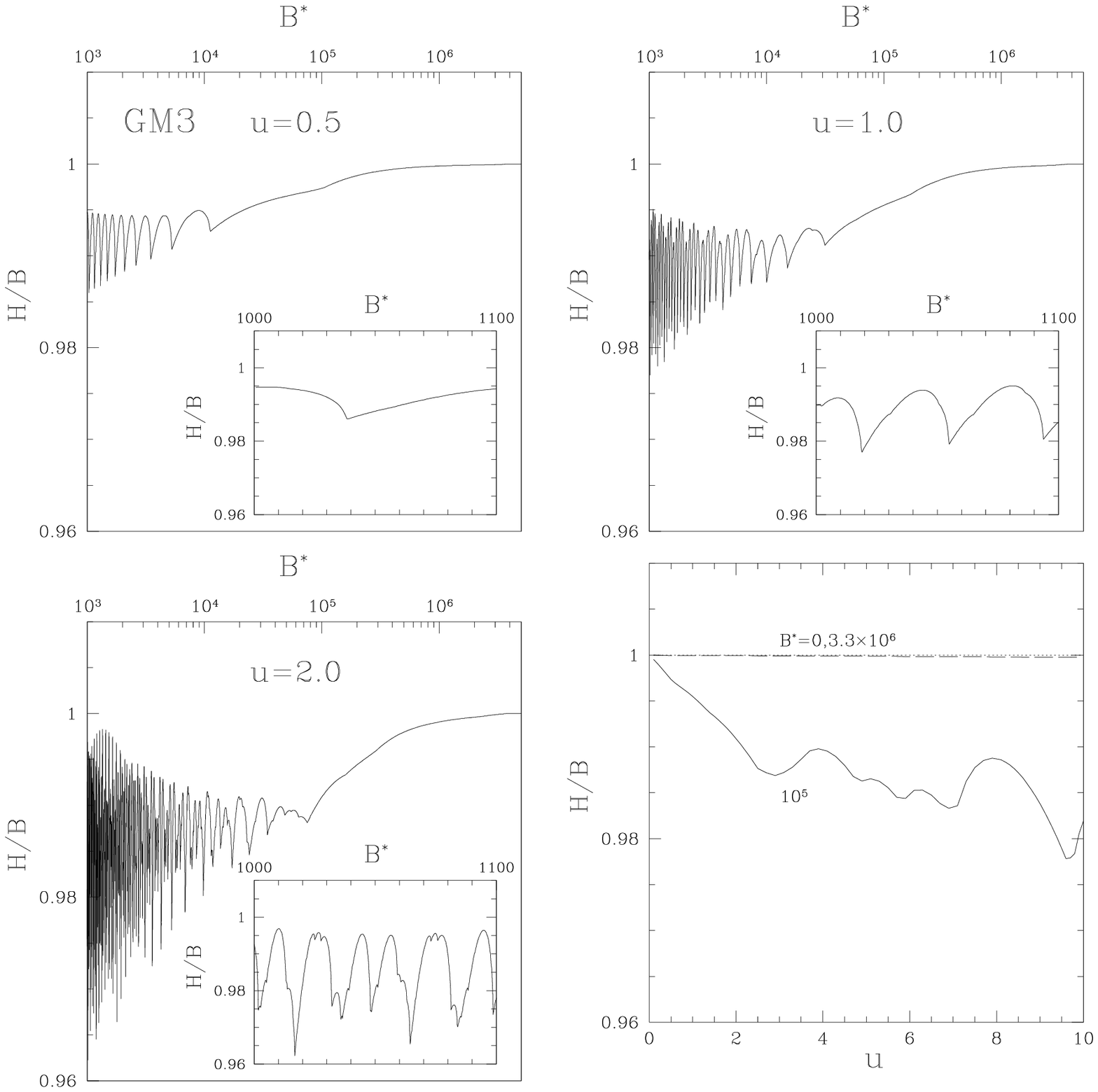}
\end{center}
\caption{}
{\label{fig6}}
\end{figure}

Figure \ref{fig6}, to be compared with Figure \ref{fig2}, shows
$H/B$ as functions of both $B^*$ and $u$ for the baseline model GM3.
The origin of the oscillations is similar to that discussed
in conjunction with Figure \ref{fig2}, but there is an overall
reduction of approximately 1\% in $H/B$ caused chiefly
by the magnetization of the neutron.

\begin{figure}
\begin{center}
\epsfxsize=6.in
\epsfysize=7.in
\epsffile{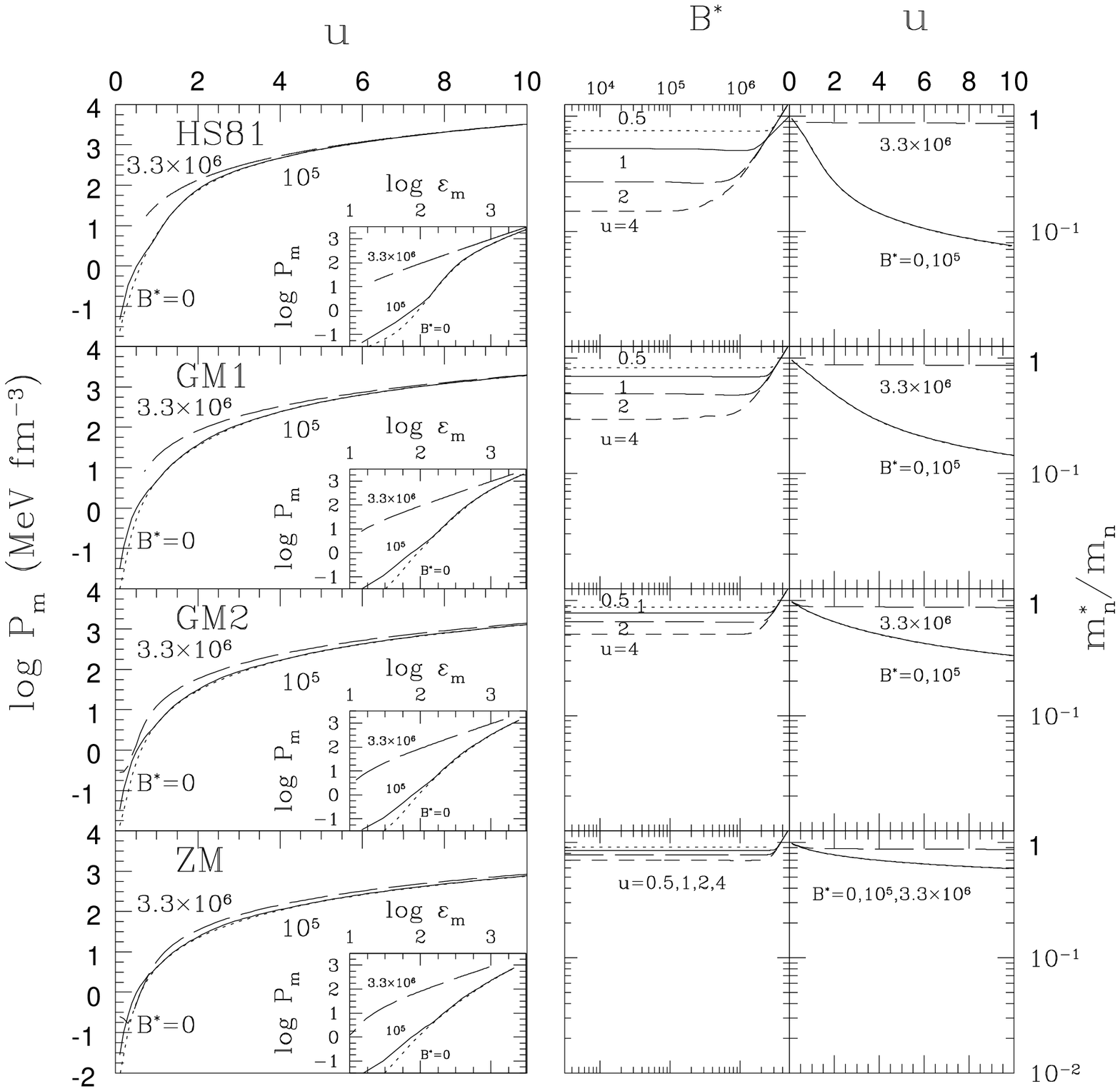}
\end{center}
\caption{}
{\label{fig7}}
\end{figure}

In Figure  \ref{fig7} (to be compared with Figure \ref{fig3}), we compare
results among the models HS81, GM1, GM2, and ZM with the
intention of extracting  generic trends induced by the inclusion of
magnetic moments.   The pressure and effective masses share the
qualitative trends exhibited by model GM3 (shown in Figure \ref{fig5}),
although quantitative
differences persist between the models.
The stiffness induced by the inclusion of magnetic moments emerges as
a general trend, and remaining differences are
principally due to variations in the underlying stiffness,
effective mass, and symmetry energies of these models..

\section{SUMMARY AND OUTLOOK}

We have developed the methodology necessary to consistently
incorporate the effects of magnetic fields on the EOS in
multicomponent, interacting matter, including a covariant description
for the inclusion of the anomalous magentic moments of nucleons.  This
methodology is necessary because in the presence of the field all
thermodynamic quantities inherit the dimensionful scale set by the
magnetic field, which necessarily affects the composition and hence
the EOS of matter.  By employing a field theoretical-apporach which
allows the study of models with different high density behaviors, we
found that the results of incorporating strong magnetic fields were
not very dependent upon the precise form of the model for the
nucleon-nucleon interaction.  The generic effects included softening
of the EOS due to Landau quantization, which is, however, overwhelmed
by stiffening due to the incorporation of the anomalous magnetic
moments of the nucleons.  These effects become significant for fields
in excess of $B^*\sim10^5$, 
for which neutrons become completely spin polarized. Note that this
field strength is
substantially less than the proton critical field.  In addition, the
inclusion of ultra-strong magnetic fields leads to a reduction in the
electron chemical potential and an increase in proton fraction.  These
compositional changes have implications for neutrino emission via the
direct Urca process and, thus, for the cooling of neutron stars.  The
magnetization of the matter never appears to become very large, as the
value of $|H/B|$ never deviates from unity by more than a few percent.
However, it remains to be seen what effects the magnetization of
matter will have on the structure and transport properties of neutron
stars.

It is worthwhile to note here that the qualitative effects of strong
magnetic fields found in the relativistic field-theoretical
description of dense matter would also be found in non-relativistic
potential models. This is because the phase space of charged particles
is similarly affected in both approaches by the presence of magnetic
fields.  The effects due to the anomalous magnetic moments would,
however, enter linearly in a non-relativistic approach (see \S4), and
would thus be more dramatic in this case.  It would be also be
instructive to study the effects of magnetic fields including
many-body correlations.

It would be useful to also consider cases in which strangeness-bearing
hyperons, a Bose (pion or kaon) condensate or quarks, are present in
dense matter.   The covariant description of the anomalous magnetic
moments developed in this work may be utilized to include hyperons, 
which are likely to be present in dense matter (Glendenning 1982, 1985;
Weber \& Weigel 1985; Kapusta \& Olive 1990;  Ellis, Kapusta \& Olive
1991;  Glendenning \& Moszkowski 1991;  Sumiyoshi \& Toki 1994;
Prakash et al. 1997 and references therein).
The anomalous magnetic moments of hyperons are mostly known. The negatively
charged hyperons,  the neutral $\Lambda$, and $\Xi^0$   all have
negative anomalous magnetic moments.  $\Sigma^+$ and $\Sigma^0$  are
the only hyperons with positive anomalous magnetic moments.    The
effects of Landau quantization on hyperons would be to soften the EOS
relative to the case in which magnetic fields are absent. However,
in the presence of strong magnetic fields, all of the hyperons will be
spin  polarized due to magnetic moment interactions with the field.
This  would cause their degeneracy pressures to increase compared to
the  field-free case.   The resultant of these two opposing
effects will depend on the relative concentrations of the various
hyperons,  which in turn  depends sensitively on the hyperon-meson
interactions for which only a  modest amount of guidance is available
(Glendenning \& Moszkowski 1991,  Knorren, Prakash \& Ellis 1995,
Schaffner \& Mishustin 1996).    For choices of $\Sigma^--$meson
interactions that favor the appearence  of $\Sigma^-$ hyperons at
relatively low densities,  the concentrations of the positively
charged particles, $p$ and $\Sigma^+$, may be expected to increase in
the  presence of strong magnetic fields.     It would thus appear that
the effects of including hyperons will not drastically alter the
qualitative trends of increasing the concentrations of positively
charged particles found in the case of $npe\mu$ matter.   
The main physical effects found in the absence of hyperons, namely 
increasing the stiffness of matter, and allowing
the direct Urca process  (Lattimer et al. 1991; Prakash et al. 1992)
to occur, probably would not change, either. Feedback effects due to
mass and energy shifts may, however, alter these expectations. Thus,
detailed calculations are required to
ascertain the influence of magnetic fields in multi-component
matter. Work on this topic is currently in progress and will be
reported separately.

It is intriguing that Bosons (pions and
kaons), which have zero magnetic moment, do not feel the magnetic fields
as fermions do.  Similarly, quarks without sub-structure also have
no anomalous magnetic moments. Thus, intense magnetic fields in
the cores of stars containing a Bose condensate or quark matter might
serve as a useful discriminant compared to those containing baryonic
matter.

Work is in progress (Cardall et al. 1999) to complete a fully
self-consistent calculation of neutron star structure including the
combined effects of the direct effects of magnetic fields on the EOS,
which we have developed in this paper, and general relativistic
structure.  The findings will help answer questions concerning the
largest frozen-in magnetic field that a stationary neutron star can
possess, and what the structure of stars with ultra-strong fields
might be.  It must be borne in mind, however, that for super-strong
fields (much higher than $B_c^p$, which is the highest field considered
in this work), the energy density in the field would be significantly
higher than the baryon mass energy density. Under such  conditions,
the internal structure of the baryons will be affected and
alternative descriptions for the EOS will become necessary.

We thank Hans Hansson for constructive suggestions concerning
the covariant description of the anomalous magnetic moments.
This work was supported in part by the NASA ATP Grant \# NAG 52863,
and by the USDOE grants DOE/DE-FG02-87ER-40317 \&
DOE/DE-FG02-88ER-40388.

\newpage

\appendix

\section{SPINORS AND ENERGY SPECTRA FOR BARYONS WITH ANOMALOUS MAGNETIC MOMENTS}

In this appendix, we derive relations for the spinors and energy spectra
for baryons with anomalous magnetic moments.  The Dirac equation is
\begin{equation}
\left[\vec{\alpha} \dotprod \left( \vec{p} - q_b \vec{A} \right)
+ \beta m_b^* + \beta \sigma_z \kappa_b B \right] \Psi_b
= E_{0,b} \Psi_b \,,
\end{equation}
where the effective momentum is given by $\vec{\pi}=\vec{p}-q_b\vec{A}$ and
$\kappa_b$ denotes the baryon anomalous magnetic moment.
The energy $E_{0,b}$ denotes the baryon energy eigenvalues when the meson
fields are absent and are related to the neutron and proton energy spectra
given in equations (\ref{ep_kap}) and (\ref{en_kap}) by
\begin{eqnarray}
E_{n,s} &=& E_{0,n} + g_{\omega_b}\omega^0 + \frac{1}{2} g_{\rho_b} \rho^0 \\
E_{p,n,s} &=& E_{0,p} + g_{\omega_b}\omega^0 - \frac{1}{2} g_{\rho_b} \rho^0 \,,
\end{eqnarray}
repectively.
Separating $\Psi_b$ in to ``big'' and ``small'' components, we obtain
\begin{eqnarray}
\left( E_{0,b} - m_b^* - \kappa_b B \sigma_z \right) \phi &=& \left( \vec{\sigma}
\dotprod \vec{\pi} \right) \chi
\label{chi} \\
\left( E_{0,b} + m_b^* + \kappa_b B \sigma_z \right) \chi &=& \left( \vec{\sigma}
\dotprod \vec{\pi} \right) \phi \,.
\end{eqnarray}
Writing $\chi$ in terms of $\phi$ (taking care to note that
the terms on the left hand side of these equations are no longer
proportional to the identity matrix because of the presence of the magnetic
moments),
equation (\ref{chi}) becomes
\begin{equation}
\left( E_{0,b} - m_b^* - \kappa_b B \sigma_z \right) \phi = \left( \vec{\sigma}
\dotprod \vec{\pi} \right) \frac{ E_{0,b} + m_b^* - \kappa_b B \sigma_z }{
\left(E_{0,b}+m_b^*\right)^2 - \left(\kappa_b B \right)^2} \left( \vec{\sigma}
\dotprod \vec{\pi} \right) \phi \,.
\end{equation}
Note that the term with $\sigma_z$ does not commute with the momentum
 operators. Therefore,
\begin{equation}
\left( E_{0,b} - m_b^* - \kappa_b B \sigma_z \right) \phi =
\frac{ E_{0,b} + m_b^* + \kappa_b B \sigma_z }{
\left(E_{0,b}+m_b^*\right)^2 - \left(\kappa_b B \right)^2} \left( \vec{\sigma}
\dotprod \vec{\pi} \right)^2 \phi  -  \frac{ 2 \kappa_b B \sigma_z}{
\left(E_{0,b}+m_b^*\right)^2 - \left(\kappa_b B \right)^2} \left( \vec{\sigma}
\dotprod \vec{\pi} \right) \pi_z \phi \,.
\end{equation}
This may be rewritten as
\begin{equation}
F_s \phi = \left( \vec{\sigma} \dotprod \vec{\pi} \right)^2 \phi
- a_s \left( \vec{\sigma} \dotprod \vec{\pi} \right) \pi_z \phi \,,
\label{gen}
\end{equation}
where $F_s$ and $a_s$ are defined as
\begin{eqnarray}
F_s &=& \left(E_{0,b}-\kappa_b B \sigma_z \right)^2 - m_b^{*~2} \nonumber \\
a_s &=& \frac{ 2 \kappa_b B \left( E_{0,b} + m_b^* - \kappa_b B \sigma_z
\right)}{\left(E_{0,b}+m_b^*\right)^2 - \left(\kappa_b B \right)^2} \,.
\end{eqnarray}
At this point it is necessary to consider individually the cases of the
protons and neutrons.
\subsubsection*{Protons}

The fact that $[ \pi_x , \pi_y ] = i \hbar (q B /c)$ suggests the
transformations
\begin{equation}
p_{\xi} = \sqrt{\frac{c}{q_p B}} \pi_x \,, \;\;
\xi=-\sqrt{\frac{c}{q_p B}} \pi_y \,.
\end{equation}
Using the identities
\begin{equation}
\left( \vec{\sigma} \dotprod \vec{a} \right)
\left( \vec{\sigma} \dotprod \vec{b} \right) =
\vec{a} \dotprod \vec{b} + i \vec{\sigma} \dotprod
\left( \vec{a} \times \vec{b} \right)
\;\;\;,\;\;\;
\vec{\pi} \times \vec{\pi} = i q_p B \hat{z}
\end{equation}
and the above transformations, equation (\ref{gen}) becomes
\begin{equation}
F_s \phi =  \left[ q_p B \left( p_{\xi}^2 + \xi^2 - \sigma_z \right) + p_z^2\right] \phi
- a_s \left[ \sqrt{\frac{c}{q_p B}} \left( \sigma_x p_xi - \sigma_y \xi \right) +
 \sigma_z p_z \right] p_z \phi \,.
\label{pdirac}
\end{equation}
The similarities between
equation (\ref{pdirac}) and that for the leptons (see, for
example,  Itzykson \& Zuber 1984),
suggests the ansatz for the spin up spinor
\begin{equation}
\phi_{+1} = e^{ik_z Z - \frac{\xi^2}{2}} \left(
\begin{array}{c}
H_n (\xi) \\
i \omega_{p,n,+1} H_{n+1} (\xi)
\end{array}
\right) \,.
\end{equation}
The two coupled differential equations for the components of $\phi$
(equations (\ref{pdirac})) reduce to two coupled
algebraic equations for the eigenvalues of $E_{0,p}$ and $\omega_{p, n, +1}$.
Explicitly,
\begin{eqnarray}
F_{+1} &=& 2 (n+1) q_p B + (1 - a_{+1})k_z^2 - a_{+1} 2(n+1) \sqrt{q_p B} k_z
\omega_{p, n, +1} \nonumber \\
F_{-1} &=& 2 (n+1) q_p B + (1 + a_{-1})k_z^2 - a_{-1} k_z \sqrt{q_p B} \omega^{-1}_{p,n,+1} \,.
\label{def_ome}
\end{eqnarray}
These may be solved to give
\begin{equation}
E_{0,p,+1} = \sqrt{k_z^2 + \left( \sqrt{m_b^{*~2} + 2 (n+1) q_p B} + \kappa_b B \right)^2} \,.
\end{equation}
With this result, it is straightforward  to solve
for $\omega_{p,n,+1}$. Lacking a
simple expression, we shall continue to refer to it as $\omega_{p,n,+1}$.
Substituting this solution for $\phi_{+1}$ into the expression
for $\chi$ gives the Dirac spinor
\begin{equation}
\Psi^p_{n,+1} = N e^{-i k_z z - \frac{\xi^2}{2}} \left(
\begin{array}{c}
H_n(\xi) \\
i \omega_{p,n,+1} H_{n+1}(\xi) \\
\frac{-k_z + 2 (n+1) \omega_{p,n,+1} \sqrt{q_p B}}{E_{0,p,+1} + m_p^* + \kappa_p B}
H_{n-1}(\xi) \\
\frac{i \sqrt{q_p B} + i \omega_{p,n,+1} k_z}{E_{0,p,+1} + m_p^* - \kappa_p B}
H_n (\xi)
\end{array} \right) \,.
\end{equation}
A similar method may be employed to find an ansatz for the spin down spinor,
\begin{equation}
\phi_{-1} = e^{ik_z Z - \frac{\xi^2}{2}} \left(
\begin{array}{c}
i \omega_{p,n,-1} H_{n-1} (\xi) \\
H_{n} (\xi)
\end{array}
\right) \,,
\end{equation}
with the energy eigenvalue
\begin{equation}
E_{0,p,-1} = \sqrt{k_z^2 + \left( \sqrt{m_b^{*~2} + 2 n q_p B} - \kappa_b B \right)^2} \,,
\end{equation}
and the Dirac spinor
\begin{equation}
\Psi^p_{n,-1} = N e^{-i k_z z - \frac{\xi^2}{2}} \left(
\begin{array}{c}
i \omega_{p,n,-1} H_{n-1} (\xi) \\
H_n(\xi) \\
\frac{-2 n i \sqrt{q_p B} - i \omega_{p,n,-1} k_z}{E_{0,p,-1} + m_p^* + \kappa_p B}
H_{n-1}(\xi) \\
\frac{k_z - \omega_{p,n,-1} \sqrt{q_p B}}{E_{0,p,-1} + m_p^* - \kappa_p B} H_n(\xi)
\end{array} \right) \,.
\end{equation}

While all quantities in this work have been calculated in the zero
temperature approximation, requiring only the postive energy spinors,
for completeness the negative energy Dirac spinors are presented below.
For the protons these may be determined
in much the same manner as that employed for the positive energy
spinors.  Defining
\begin{eqnarray}
F_s^- &=& \left( E_{0,p} + \kappa_p B \sigma_z \right)^2 - m_p^{*~2} \\
a_s^- &=& -\frac{2 \kappa_p B
\left( E_{0,p} - m^*_p + \kappa_p B \sigma_z \right)}{
\left(E_{0,p} - m^*_p \right)^2 - \left(\kappa_p B\right)^2} \,,
\end{eqnarray}
the equation for $\chi$ takes the same form as equation (\ref{gen}) where
$F_s^-$ and $a_s^-$ replace $F_s$ and $a_s$ respectively.  As a
result, precisely the same formalism employed to determine the positive
energy spinors may be used to determine the negative energy spinors.
The Dirac spinor corresponding to the energy eigenvalue
\begin{equation}
E_{0,p,+1}^- = - \sqrt{k_z^2 + \left(\sqrt{m_p^{*~2}
+ 2 \left( n + 1 \right) q_p B } + \kappa_p B \right)^2} \,,
\end{equation}
is given by
\begin{equation}
\Psi_{n,+1}^{p,-} = N e^{-i k_z Z - \frac{\xi^2}{2}}
\left(
\begin{array}{c}
\frac{-k_z + 2(n+1) \omega_{p,n,+1}^- \sqrt{q_p B}}
{E_{0,p,+1}^- - m_p^* - \kappa_p B} H_{n} (\xi) \\
\frac{i\sqrt{q_p B} + i \omega_{p,n,+1}^- k_z}
{ E_{0,p,+1}^- - m_p^* + \kappa_p B} H_{n+1} (\xi) \\
H_n(\xi) \\
i \omega_{p,n,+1}^- H_{n+1}(\xi)
\end{array}
\right) \,,
\end{equation}
where $\omega_{p,n,+1}^-$ is defined by replacing $F_s$ and $a_s$ in equations
(\ref{def_ome}).  Similarly, the Dirac spinor corresponding to the energy
eigenvalue
\begin{equation}
E_{0,p,-1}^- = - \sqrt{k_z^2 + \left(\sqrt{m_p^{*~2}
+ 2 n q_p B } - \kappa_p B \right)^2} \,,
\end{equation}
is given by
\begin{equation}
\Psi_{n,-1}^{p,-} = N e^{-i k_z Z - \frac{\xi^2}{2}}
\left(
\begin{array}{c}
\frac{-2 i n \sqrt{q_p B} - i \omega_{p,n,-1}^- k_z }
{E_{0,p,+1}^- - m_p^* - \kappa_p B} H_{n-1} (\xi) \\
\frac{k_z -  \omega_{p,n,-1}^- \sqrt{q_p B}}
{ E_{0,p,+1}^- - m_p^* + \kappa_p B} H_{n} (\xi) \\
i \omega_{p,n,-1}^- H_{n-1}(\xi) \\
H_n(\xi)
\end{array}
\right) \,.
\end{equation}

\subsubsection*{Neutrons}

In this case, the trial wave function has the same form as the
free particle solutions with unknown
coefficients, which may be determined in a manner
analougous to that employed
for the protons.  Define
\begin{equation}
G_s=F_s-k^2 + \sigma_z a_s k_z^2 \,.
\label{def_Gs}
\end{equation}
Then, equation (\ref{gen}) becomes
\begin{equation}
G_s \phi = - a_s ( \sigma_x k_x + \sigma_y k_y ) k_z \phi \,.
\label{ndirac}
\end{equation}
Note that both $G_s$ and $a_s$ are diagonal and therefore the off-diagonal terms have
been isolated on the right-hand side of equation (\ref{ndirac}).  The
similarities with the case in which $\kappa_n=0$, namely
the quadratic nature
of the momentum operators, suggests the form
\begin{equation}
\phi=e^{-i k^\mu x_\mu} \left( \begin{array}{c}
u \\
v
\end{array}
\right) \,.
\label{nansatz}
\end{equation}
Using equation (\ref{nansatz}), we obtain the coupled algebraic equations
\begin{eqnarray}
G_{+1} u &=& - a_{+1} ( k_x - i k_y ) k_z v \nonumber \\
G_{-1} v &=& - a_{-1} ( k_x + i k_y ) k_z u \,.
\label{ncomb}
\end{eqnarray}
Combining these gives
\begin{equation}
G_{+1} G_{-1} = a_{+1} a_{-1} ( k_x^2 + k_y^2 ) k_z^2 \,,
\end{equation}
which may be solved for the energy eigenvalue
\begin{equation}
E_{0,n,s} = \sqrt{ k_z^2 + \left( \sqrt{ m_n^{*~2} + k_x^2 + k_y^2}
+ s \kappa_n B \right)^2} \,.
\end{equation}
The eigenvectors may be determined, up to a normalization, by setting
\begin{eqnarray}
u &=& 1 \;\; \rightarrow \;\; v=-\frac{a_{-1}}{G_{-1}} (k_x + i k_y) k_z \\
v &=& 1  \;\; \rightarrow \;\; u = -\frac{a_{+1}}{G_{+1}} (k_x + i k_y) k_z
\end{eqnarray}
in equation (\ref{ncomb}).
It is clear from direct substitution that the first gives
the $s=+1$ and the second the $s=-1$ spinors.
Inserting these into equation
(\ref{nansatz}) and then into the expression for $\chi$ gives the
neutron Dirac spinors
\begin{equation}
\Psi^n_{+1} = N e^{-i k^{\mu} x_{\mu}} \left( \begin{array}{c}
 1 \\
 - \frac{a_{-1}}{G_{-1}} \left( k_x + i k_y \right) k_z \\
\frac{ \left[ 1-\frac{a_{-1}}{G_{-1}} \left(k_x^2 + k_y^2 \right) \right] }
{E_{0,n,+1} + M^*_n + \kappa_n B} \\
\frac{ \left[ 1 - \frac{a_-}{b_-} k_z^2 \right] \left( k_x + i k_y \right)}{
E_{0,n,+1} + M^*_n - \kappa_n B} \,,
\end{array} \right)
\end{equation}
\begin{equation}
\Psi^n_{-1} = N e^{-i k^{\mu} x_{\mu}} \left( \begin{array}{c}
 - \frac{a_{+1}}{G_{+1}} \left( k_x - i k_y \right) k_z \\
 1 \\
\frac{ \left[ 1 - \frac{a_{+1}}{G_{+1}} k_z^2 \right] \left( k_x - i k_y \right)}{
E_{0,n,-1} + M^*_n + \kappa_n B} \\
-\frac{ \left[ 1 + \frac{a_{+1}}{G_{+1}} \left(k_x^2 + k_y^2 \right) \right] k_x}
{E_{0,n,-1} + M^*_n - \kappa_n B}
\end{array} \right) \,.
\end{equation}

In order to determine the negative energy Dirac spinors for the neutron, an
approach analogous to that employed in determining the negative energy
Dirac spinors for the protons may be used.  Define $G_s^-$
by replacing $F_s$ and $a_s$ by $F_s^-$ and $a_s^-$, respectively,
in equation (\ref{def_Gs}).  As in the case of the protons, this produces an
equation
for $\chi$ which is of the same form as that employed for $\phi$ in the
derivation of the positive energy spinors.  Proceeding in the same manner
as before, one finds that the Dirac spinors corresponding to the
energy eigenvalues
\begin{equation}
E_{0,n,s}^- = -\sqrt{k_z^2 + \left( \sqrt{m_n^* + k_x^2 + k_y^2}
+ s \kappa_n B \right)^2} \,,
\end{equation}
are given by
\begin{eqnarray}
\Psi_{+1}^{n,-} &=& N e^{-i k^\mu x_\mu} \left(
\begin{array}{c}
\frac{\left[ 1 - \frac{a_{-1}^-}{G_{-1}^-} \left( k_x^2 + k_y^2 \right)
\right] k_z}{E_{0,n,+1} - m_n^* - \kappa_n B} \\
\frac{\left[ 1 - \frac{a_{-1}^-}{G_{-1}^-} k_z^2 \right]
\left( k_x + i k_y \right) }{E_{0,n,+1} - m_n^* + \kappa_n B} \\
1 \\
-\frac{a_{-1}^-}{G_{-1}^-} \left( k_x + i k_y \right) k_z
\end{array}
\right)
\\
\Psi_{-1}^{n,-} &=& N e^{-i k^\mu x_\mu} \left(
\begin{array}{c}
\frac{\left[ 1 - \frac{a_{+1}^-}{G_{+1}^-} k_z^2 \right]
\left( k_x - i k_y \right) }{E_{0,n,-1} - m_n^* - \kappa_n B} \\
-\frac{\left[ 1 + \frac{a_{+1}^-}{G_{+1}^-} \left( k_x^2 + k_y^2 \right)
\right] k_z}{E_{0,n,-1} - m_n^* + \kappa_n B} \\
-\frac{a_{+1}^-}{G_{+1}^-} \left( k_x - i k_y \right) k_z \\
1
\end{array}
\right) \,.
\end{eqnarray}

\newpage


\newpage


\begin{center}
\centerline{TABLE 1}
\vspace*{0.15in}
\centerline{NUCLEON-MESON COUPLING CONSTANTS}
\begin{tabular}{ccccccccccccr}
\hline
Model & $n_s$ & $-B/A$ & $M^*/M$ & $K_0$ & $a_{\mbox{\tiny sym}}$ &
$g_{\sigma_N}/m_\sigma $
& $g_{\omega_N}/m_\omega$ & $g_{\rho_N}/m_\rho$ & $b$ & $c$ \\
\hline
 HS81 & 0.148 & 15.75 & 0.54 & 545 & 35.0 & 3.974 & 3.477 & 2.069 & 0.0 & 0.0 \\
\hline
 GM1 & 0.153 & 16.30 & 0.70 & 300 & 32.5 & 3.434 & 2.674 & 2.100 & 0.002947 &
 $-0.001070$ \\
 GM2 & 0.153 & 16.30 & 0.78 & 300 & 32.5 & 3.025 & 2.195 & 2.189 & 0.003478 &
 0.01328 \\
 GM3 & 0.153 & 16.30 & 0.78 & 240 & 32.5 & 3.151 & 2.195 & 2.189 & 0.008659 &
 $-0.002421$ \\
\hline
 ZM  & 0.160 & 16.00 & 0.86 & 225 & 32.5 & 2.736 & 1.617 & 2.185 & 0.0 & 0.0 \\
\hline
\hline
\end{tabular}
\end{center}
\vspace*{0.15in} NOTE.-- Coupling constants for the HS (Horowitz \&
Serot 1981), GM1-3 (Glendenning \& Moszkowski  1991), and ZM (Zimanyi
\& Moszkowski 1990; 1992) models.  The couplings are chosen to
reproduce the binding energy  $B/A$ (MeV), the nuclear saturation
density $n_s$ ($\mbox{fm}^{-3}$),  the Dirac effective mass $M^*$ in
units of the baryon mass $M$, and the the symmetry energy
$a_{\mbox{\tiny sym}} (MeV)$. The nuclear matter compression modulus $K_0$
(MeV) for the different  models are also listed.

\newpage

\section*{FIGURE CAPTIONS}

\ni FIG. 1.-- Matter pressure $P_m$, nucleon Dirac effective mass
$m_n^*/m_n$, and concentrations $Y_i=n_i/n_b$ as functions of the
density $u=n_b/n_s$ (left panels; $n_s=0.16~{\rm fm}^{-3}$ is the
fiducial nuclear saturation density) and magnetic field strength
$B^*=B/B_e^c$ (right panels; $B_e^c=4.414 \times 10^{13}$ Gauss is the
electron critical field), for the model GM3.  The inset in the upper
left panel shows $P_m$ as a function of the matter energy density
$\varepsilon_m$.  The curve labeled $P_f$ in the upper right panel
shows the $B^2/8\pi$ contribution to the total pressure.  The inset in
the lower left panel shows the effective mass as a function of $B^*$.
In the lower right panel, the electron and neutron concentrations have
been suppressed for clarity ($Y_e=Y_p-Y_\mu$ and $Y_n=1-Y_p$).
\vspace*{0.5cm}

\ni FIG. 2.-- The ratio of the induced to applied magnetic field $H/B$
as functions of the density and magnetic field strength, for
the model GM3.  The insets show $H/B$ in expanded scales to highlight
the effects of including several components.
\vspace*{0.5cm}

\ni FIG. 3.-- Matter pressure $P_m$ and the nucleon Dirac effective
mass $m_n^*/m_n$ for the models shown in Table 1 (with the exception
of model GM3, whose results are displayed in Figures 1 and 2), as
functions of the density and magnetic field strength.  The insets in
the left panels show $P_m$ as a function of the matter energy density
$\varepsilon_m$.
\vspace*{0.5cm}

\ni FIG. 4.-- Matter pressure $P_m$ and concentrations $Y_i=n_i/n_b$
as functions of the density and magnetic field strength for a charge
neutral, beta-equilibrated, non-interacting $npe\mu$ gas with and
without the inclusion of the nucleon anomalous magnetic moments
$\kappa_b$.  The curve labeled $P_f$ in the upper right panel shows
the $B^2/8\pi$ contribution to the total pressure.  The lower left
panel shows the enhancement in the pressure, as a function of energy
density, due to the presence of magnetic fields.  In the lower right
panel, the electron and neutron concentrations have been suppressed
for clarity ($Y_e=Y_p-Y_\mu$ and $Y_n=1-Y_p$).
\vspace*{0.5cm}

\ni FIG. 5.-- Same as Figure 1, except that the nucleon anomalous
magnetic moments are now included.
\vspace*{0.5cm}

\ni FIG. 6.-- Same as Figure 2, except that the nucleon anomalous
magnetic moments are now included.
\vspace*{0.5cm}

\ni FIG. 7.-- Same as Figure 3, except that the nucleon anomalous
magnetic moments are now included.
\vspace*{0.5cm}

\newpage

\begin{thebibliography}{99}

\bibitem {as} Abrahams, A. M., \& Shapiro, S. L. 1991, ApJ, 374, 652

\bibitem{bb} Boguta, J. \& Bodmer, A. R. 1977, Nucl.\ Phys.,\ A292, 413

\bibitem{BD:rqm} Bjorken, J., \&  Drell, S. 1964,
in Relativistic Quantum Mechanics (McGraw-Hill: New York)

\bibitem{BH:magsus} Blandford, R.D. \& Hernquist, L.,
1982, J. Phys. C: Solid State Phys., 15, 6233

\bibitem{boc} Boucquet, M., Bonozzola, S., Gourgoulhon, E., \& Novak, J.
1995 Astron. \& Astrophys. 301, 757


\bibitem{cv}Canuto, V., \& Ventura, J.  1977, Fund. Cosmic Phys., 2, 203

\bibitem{Card} Cardall, C.Y., Broderick, A., Prakash, M. \& Lattimer, J.M.
1999, to be published

\bibitem{Ch0} Chakrabarty, S. 1996, Phys. Rev. D54, 1306

\bibitem{ch:dense} Chakrabarty, S., Bandyopadhyay, D., \& Pal, S. 1997
Phys. Rev. Lett. 78, 2898

\bibitem{dn0} Duncan, R.C. \& Thompson, C. 1992, ApJ, 392, L9

\bibitem{dn1} Duncan, R.C. \& Thompson, C. 1996, ApJ, 469, 764

\bibitem{eko} Ellis, J., Kapusta, J. \& Olive, K. A. 1991, 
Nucl.\ Phys.\ B348, 345


\bibitem{fgp}Fushiki, I., Gudmundsson, E.H., \& Pethick, C. J.
1989, ApJ, 342, 958

\bibitem{fgpy}Fushiki, I., Gudmundsson, E.H.,  Pethick, C. J., \&
Yngvason, J. 1992, Ann. Phys., 216, 29


\bibitem{gm1} Glendenning, N. K. 1982, Phys.\ Lett.\ 114B, 392

\bibitem{gm2} Glendenning, N. K. 1985, Astrophys.\ J.\ 293, 470



\bibitem{GM} Glendenning, N.K. \& Moszkowski, S.A. 1991, Phys. Rev. Lett.,
67, 2414


\bibitem{hs} Horowitz, C.J., \& Serot, B.D. 1981, Nucl. Phys. A368, 503

\bibitem{iz} Itzykson, C., \& Zuber, J.-B. 1980, Quantum Field Theory,
(McGraw-Hill: New York)

\bibitem{LJ} Johnson, M.H \& Lippman, B.A. 1950, Phys. Rev. 77, 702


\bibitem{ko} Kapusta, J. \& Olive, K. A. 1990, Phys.\ Rev.\ Lett.\ 64, 13

\bibitem{kpe} Knorren, R., Prakash, M. \& Ellis, P.J. 1995, 
Phys. Rev., C52, 1855

\bibitem{k} Kouveliotou. C. et al., 1998, Nature, 393, 235

\bibitem{ls} Lai, D. \& Shapiro, S. 1991, ApJ, 383, 745

\bibitem{llp} Landau, L.D., Lifshitz, E.M., \& Pitaevski\u{i}, L.P. 1984,
Electrodynamics of Continuous Media 2nd ed. (New York: Pergamon)

\bibitem{lpph} Lattimer, J.M., Pethick, C.J., Prakash, M. \& Haensel, P. 1991,
Phys. Rev. Lett. 66, 2701  

\bibitem{Mel}Melatos, A. 1999, ApJ, 519, L77

\bibitem{MS:nucl} M\"uller, H., \& Serot, B.D. 1996, Nucl. Phys., A606, 508

\bibitem{pac}Paczynski, B. 1992, Acta Astron., 42, 145


\bibitem{pr:eos} Prakash, M., Bombaci, I., Manju Prakash, Ellis, P.J.,
Lattimer, J.M., \& Knorren, R. 1997, Phys. Rep. 280, 1

\bibitem{pplp} Prakash, M., Prakash Manju, Lattimer, J.M. \& Pethick, C.J. 
1992, ApJ, 390, L77

\bibitem{rfgpy} R\"{o}gnvaldsson, Fushiki, I.,
Gudmundsson, E. H., C. J. Pethick, \& \"O. E., Yngvason, J.,
ApJ, 416, 276.

\bibitem{sm} Schaffner, J. \& Mishustin, I.N. 1996, Phys. Rev., C53, 1416

\bibitem{st} Sumiyoshi, K. \& Toki, H. 1994, Astrophys.\ J.\ 422, 700

\bibitem{th0} Thompson, C. \& Duncan, R.C. 1995, MNRAS, 275, 255

\bibitem{th1} Thompson, C. \& Duncan, R.C. 1996, ApJ, 473, 322

\bibitem {tryg}Thorlofsson, A.,
R\"{o}gnvaldsson, \"O. E., Yngvason, J., \& Gudmundsson E. H. 1998,
ApJ, 502, 847


\bibitem{ww} Weber, F. \&  Weigel, M. K. 1989, Nucl.\ Phys.,\ A505, 779

\bibitem{yz} Y. F. Yuan, Y. F.  \& Zhang, J. L. 1999, 
ApJ, 525, 920 

\bibitem{zm} Zimanyi, J. \& Moszkowski, S.A. 1990, Phys. Rev. C42, 1416; \\
------------- 1992, Phys. Rev. C45, 844

\end{thebibliography}
\end{document}